\documentclass[a4paper,fleqn,usenatbib]{mnras}

\usepackage[T1]{fontenc}
\usepackage{ae,aecompl}
\usepackage{adjustbox}
\usepackage{graphicx}	% Including figure files
\usepackage{amsmath}	% Advanced maths commands
\usepackage{amssymb}	% Extra maths symbols

\usepackage{color}
\usepackage{float}
\usepackage{txfonts}
\usepackage{units}
\usepackage{url}
\usepackage{slantsc}

\newcommand{\mhcm}{\rm m_{H}~cm^{-3}}
\newcommand{\msunyr}{{\rm M_{\odot} yr^{-1}}}
\newcommand{\cmc}{\rm cm^{-3}}
\newcommand{\vsc}{v_{\rm s,c}}
\newcommand{\vw}{v_{\rm w}}  
\newcommand{\kms}{{\rm km~s^{-1}}}  
\newcommand{\rhoc}{\rho_{\rm c}}
\newcommand{\rhow}{\rho_{\rm w}}
\newcommand{\tc}{t_{\rm cool}}
\newcommand{\cv}{C_{\rm v}}
\newcommand{\Lkin}{L_{\rm kin}}
\newcommand{\Lkinf}{L_{\rm kin,45}}
\newcommand{\Rc}{R_{\rm c}} 
\newcommand{\ergs}{{\rm erg~s^{-1}}}
\newcommand{\ergcm}{{\rm erg~cm^{-3}}}
\newcommand{\nh}{n_{\rm H}}
\newcommand{\Bd}{B_{\rm d}}
\newcommand{\Emax}{E_{\rm max}}
\newcommand{\Emaxp}{E_{\rm max,p}}
\newcommand{\Emaxe}{E_{\rm max,e}}
\newcommand{\Ecr}{E_{\rm cr}}

\newcommand{\Lgm}{L_{\gamma}}
\newcommand{\Lagn}{L_{\rm AGN}}
\newcommand{\kbow}{k_{\rm bow}}
\newcommand{\Ncr}{N_{\rm cr}}
\newcommand{\Np}{N_{\rm p}}
\newcommand{\Ne}{N_{\rm e}}

\newcommand{\Egm}{E_{\gamma}}
\newcommand{\Kep}{K_{ep}}

\newcommand{\Kinj}{K_{\rm inj}}
\newcommand{\Eep}{\eta_{ep}}
\newcommand{\ecrp}{\epsilon_{\rm crp}}
\newcommand{\ecre}{\epsilon_{\rm cre}}
\newcommand{\tdyn}{t_{\rm dyn}}
\newcommand{\nuc}{\nu_c}
\newcommand{\GeV}{{\rm GeV}}
\newcommand{\TeV}{{\rm TeV}}
\newcommand{\Gm}{\Gamma}
\newcommand{\epm}{e^{\pm}}
\newcommand{\nue}{\nu_e}
\newcommand{\anue}{\bar{\nu}_e}
\newcommand{\num}{\nu_{\mu}}
\newcommand{\anum}{\bar{\nu}_{\mu}}
\newcommand{\be}{\begin{equation}}
\newcommand{\ee}{\end{equation}}

\title[Gamma-ray and Radio Afterglows]{Years Delayed Gamma-ray and Radio Afterglows Originated from TDE Wind-Torus Interactions}

\author[Mou \& Wang]{Guobin~Mou$^{1,2,3,4}$\thanks{E-mail: gbmou@whu.edu.cn}
and Wei~Wang$^{1,2}$\thanks{E-mail: wangwei2017@whu.edu.cn}
\\
$^{1}$School of Physics and Technology, Wuhan University, Wuhan 430072, China  \\
$^{2}$WHU-NAOC Joint Center for Astronomy, Wuhan University, Wuhan 430072, China \\
$^{3}$CAS Key Laboratory for Research in Galaxies and Cosmology, Department of Astronomy, University of Science and Technology of China, Hefei 230026, China  \\
$^{4}$School of Astronomy and Space Science, University of Science and Technology of China, Hefei 230026, China \\
}

\date{in original form 2020 Dec 05}

\pubyear{ -- }

\begin{document}
\label{firstpage}
\pagerange{\pageref{firstpage}--\pageref{lastpage}}
\maketitle

% Abstract of the paper
\begin{abstract} 
Tidal disruption events (TDEs) that occur in active galactic nuclei (AGN) with dusty tori are a special class of sources. 
TDEs can generate ultrafast and large opening-angle wind, which will almost inevitably collide with the preexisting AGN dusty tori a few years later after the TDE outburst. 
The wind-torus interactions drive two kinds of shocks: the bow shocks at the windward side of the torus clouds, and the cloud shocks inside the torus clouds. 
In a previous work, we proved that the shocked clouds will give rise to considerable X-ray emissions which can reach $10^{41-42}~\ergs$ (so called \emph{years delayed X-ray afterglows}).   
In this work, we focus on the radiations of high energy particles accelerated at both shocks. 
Benefitting from the strong radiation field at the inner edge of the torus, the inverse Compton scatterings of AGN photons by relativistic electrons at bow shocks dominate the overall gamma-ray radiation. 
The gamma-ray luminosity can reach $10^{41}~\ergs (L_{\rm kin}/10^{45}{\rm erg s^{-1}})$, where $L_{\rm kin}$ is the kinetic luminosity of TDE wind. 
Synchrotron radiation at bow shocks contributes to the radio afterglow with a luminosity of 10$^{38-39} ~\ergs (L_{\rm kin}/10^{45}{\rm erg s^{-1}})$ at 1-10 GHz if the magnetic field is 100 mGauss, and extends to infrared with a luminosity of $\sim 10^{39-40}~{\rm erg s^{-1}} (L_{\rm kin}/10^{45}{\rm erg s^{-1}})$. 
Our scenario provides a prediction of the years delayed afterglows in  multiple wavebands for TDEs and reveals their connections. 
\end{abstract} 


\section{Introduction}

When a star occasionally plunges into the tidal radius of supermassive back hole (SMBH), it will be disrupted and give rise to a tidal disruption event (TDE; \citealt{hills1975, rees1988}). The bound debris will fall back to the SMBH and generate luminous outburst in optical/UV or X-ray band which declines on the timescale of months to years (e.g., \citealt{komossa2015,van2019}).
If the pericenter of the star is very close to the black hole (a few Schwarzschild radius), the relativistic apsidal precession will be strong. After passing the pericenter, the falling debris soon collides with the still in-falling stream (self-crossing) at a very high relative speed. This violent collision can also generate wind, of which the kinetic energy can reach up to $10^{51-52}$erg (\citealt{lu2020}) with mean speed of $0.01-0.1$c (\citealt{sadowski2016}, see also \citealt{jiangyf2016}). On the other hand, the circularized bound debris will trigger a short-term high accretion rate and generate strong wind, of which the kinetic luminosity is $10^{44-46}\ergs$(\citealt{dai2018, curd2019}).   
Observationally, the existence of TDE wind can be confirmed directly in UV and X-ray band (e.g., \citealt{blanchard2017, blagorodnova2019, nicholl2020, hung2019}), and the high kinetic energy of wind has been indirectly inferred by radio emissions for some TDE candidates (e.g., \citealt{coppejans2020, alexander2020}).  
In this paper, we refer to the ``wind'' as a synonym for large opening-angle ``outflows'', relative to the collimated jet. 
When a TDE occurs in an AGN with a dusty torus which is composed of amounts of clouds (\citealt{elitzur2012, netzer2015}), the transient strong radiation and energetic wind will trigger transient echoes or afterglows.  
Depending on the distance of the torus, typically tens of days later after the primary outburst, the optical/UV/soft X-ray photons from TDE outburst irradiate the surrounding dusty clouds and result in an infrared echo (\citealt{lu2016, dou2016, jiang2016, van2016}; 
 see also \citealt{mattila2018} for an infrared echo by polar dust ). A few years later, the fast and strong TDE wind impacts the torus violently and drive cloud shocks inside the clouds. The cloud materials swept by cloud shock will radiate in X-rays (``X-ray afterglow'', \citealt{jiang2019, mou2021}). Depending on the strength of the TDE wind and the physics of the torus, the X-ray luminosity can be as high as $10^{41-42}~\ergs$, of which the X-ray properties can be used to constrain the physics of the TDE wind and the torus (\citealt{mou2021}). 

\begin{figure*}
\includegraphics[width=1.7\columnwidth]{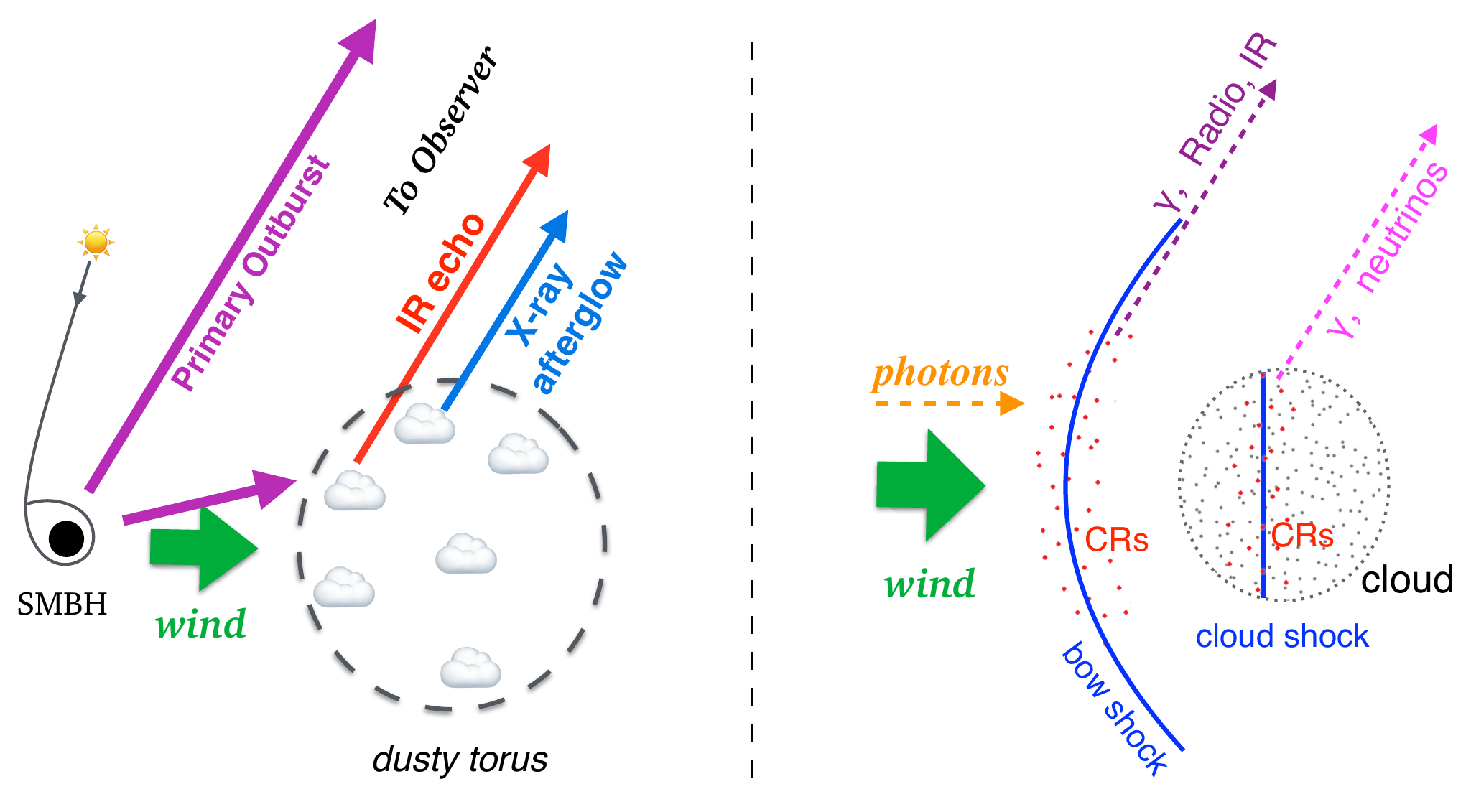}
 \caption{Sketch of the interactions between the TDE wind and AGN torus. The left panel shows the overall processes, including the primary outburst in optical/UV/soft X-ray band due to a sudden increased accretion rate, the subsequent infrared echo with a time lag of a few tens of days which is caused by reproduces of the irradiated dusty torus. A few years later, the TDE wind reaches the torus, and produces an X-ray afterglow by shocking the clouds. The right panel shows an enlarged view of a single cloud undergoing wind-cloud interactions. The red dots mark the cosmic rays accelerated at the bow shock and the cloud shock. Those CRs will produce 
{\bf afterglows of gamma-ray, radio-infrared and neutrino. }  } 
 \label{fig1}
\end{figure*}

The collision between the TDE wind and torus will not only produce X-rays, but also accelerate charged particles. 
The collision leads to two kinds of shocks: the bow shock at the windward side of the cloud, and the cloud shock inside the cloud (\citealt{mckee1975}). According to the diffuse shock acceleration theory (DSA, e.g., see \citealt{drury1983} for reviews), the fast shock is able to accelerate the particles to relativistic ones. 
At the bow shock, as exposed in the circumstance of a strong radiation field in the vicinity of an AGN, the cosmic ray electrons (CRe) will efficiently produce gamma-rays via 
 inverse Compton scattering (ICS) 
of AGN photons, and radio emissions via synchrotron radiation.  
At the cloud shock in the dense cloud, the CRp will undergo proton-proton collisions (pp collisions) and generate gamma-rays and neutrinos.   
Thus, it is naturally expected that there would be gamma-ray/radio, or even neutrino afterglows accompanied with the X-ray afterglow (see Figure \ref{fig1}). 
Currently, there are very few TDE candidates with gamma-ray or neutrino detections (e.g., see \citealt{stein2021} for a recent report of a high energy muon neutrino event associated with a TDE candidate).  
Gamma-rays/neutrinos of TDE are mainly at the stage of theoretical studies. 
\citet{cheng2007} proposed that if the accretion power ($\dot M c^2$) can be converted into the jet power very efficiently ($\sim 10\%$),  taking the TDE rate of $10^{-5}$ yr$^{-1}$, the pp collisions in the galactic center environment are able to sustain a gamma-ray emission of $\sim 10^{38}~\ergs$ in which the peak gamma-ray luminosity can reach $10^{40}~\ergs$. 
\citet{chen2016} investigated the interaction of unbound debris stream (the half of the stellar materials that gaining positive mechanical energy by tidal force) and the dense molecular clouds, and estimated that the gamma-ray afterglow will arise hundreds of years after the TDE with a maximum gamma-ray luminosity of $\sim 1\times 10^{39}~\ergs$. 
 \citet{murase2020} investigated the neutrino and gamma-ray emission that originated from the inner corona/disk region, and the disk wind or stream-stream interaction region, and argued that these processes should be accompanied by soft gamma-rays as well as optical/UV emission.  
Recently, \citet{liu2020} studied the interactions of CRp accelerated in jet and a very intense radiation field inside the optical/UV photosphere of TDE via P$\gamma$ reactions, which can lead to a gamma-ray intrinsic radiation of $\sim 10^{42}~\ergs$.   
In radio band, there are about ten TDEs detected with radio emissions (see review by \citealt{alexander2020}). 
The peak radio luminosities are $10^{36-42}\ergs$, with time lags spanning from days to years. Such radio emissions are thought to be generated by synchrotron emissions of relativistic electrons, which are accelerated in forward/external shocks driven by wind in the diffuse ISM, or reverse/internal shocks driven by jets. 

Here, for TDEs occurring in AGN, we continue the study on the interactions of TDE wind and torus, and predict years delayed afterglows in multiple wavebands. Combining with multi-band afterglows, a physical model can be more reliably supported or denied. 

The probability of TDE occurring in AGN is not quite low. 
\citet{elitzur2006} argued that the torus should disappear in low-luminosity AGN when the bolometric luminosities are below $\sim 10^{42}~\ergs$. This corresponds to the Eddington ratio of $10^{-2}-10^{-3}$ for $10^{6-7}$ solar mass BH. 
The duty cycle (the fraction of SMBH lifetime in the active phase) above this Eddington ratio is $\sim 10^{-1}$ (e.g., \citealt{gan2014}). 
Thus, we argue that the 
proportion of TDEs occurring in AGN with torus should be in the order of 10\% of the overall TDEs. 
More generally, if we relax the condition to \emph{TDEs occurring in SMBHs surrounded by clouds} while regardless of whether the SMBH is active or not (e.g., within $\sim$1 pc from Sgr A* exist mini-spiral and circumnuclear disk, \citealt{martin2012}), the proportion 
will be higher. 

The rest of the paper is organized as follows. In Section 2 we briefly introduce the TDE wind and torus interactions and the environment for particle acceleration. In Section 3 we describe the properties of CRp and CRe involving cooling. We present the results of gamma-ray, neutrino and radio emissions in Section 4, and give a summarize and discussion in Sections 5. Details on adiabatic cooling of CRs and the calculations or analysis of gamma-ray, neutrino and synchrotron emissions are gathered in appendix. 

%%%%%%%%%%%%%%%%%%%%%%%%%%%%%%%%%%
%%%%%%%%%%%%%%%%%%%%%%%%%%%%%%%%%%

\begin{figure}
\includegraphics[width=1.0\columnwidth]{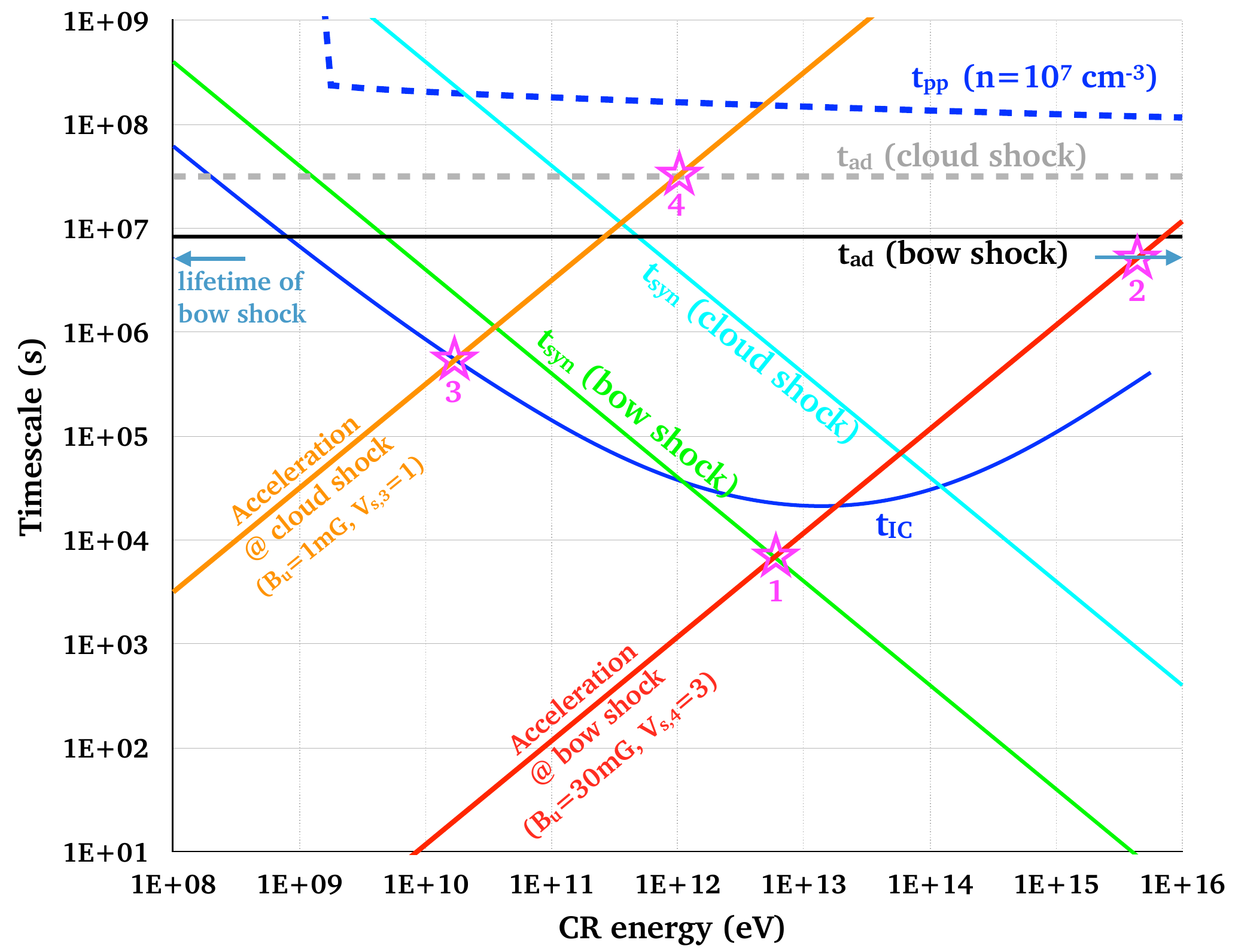}
 \caption{Timescales of various processes as functions of CR energy. As marked in magenta stars, the intersections of the acceleration timescale lines and the energy loss timescale/shock lifetime lines provide the maximum energy of CR: 
 1) $\sim$ TeV for CRe at bow shock, 
 2) PeV for CRp at bow shock, 
 3) 10 GeV for CRe at cloud shock, 
 4) 1 TeV for CRp at cloud shock. }   
 \label{fig2}
\end{figure} 

\begin{figure*}
\includegraphics[width=1.4\columnwidth]{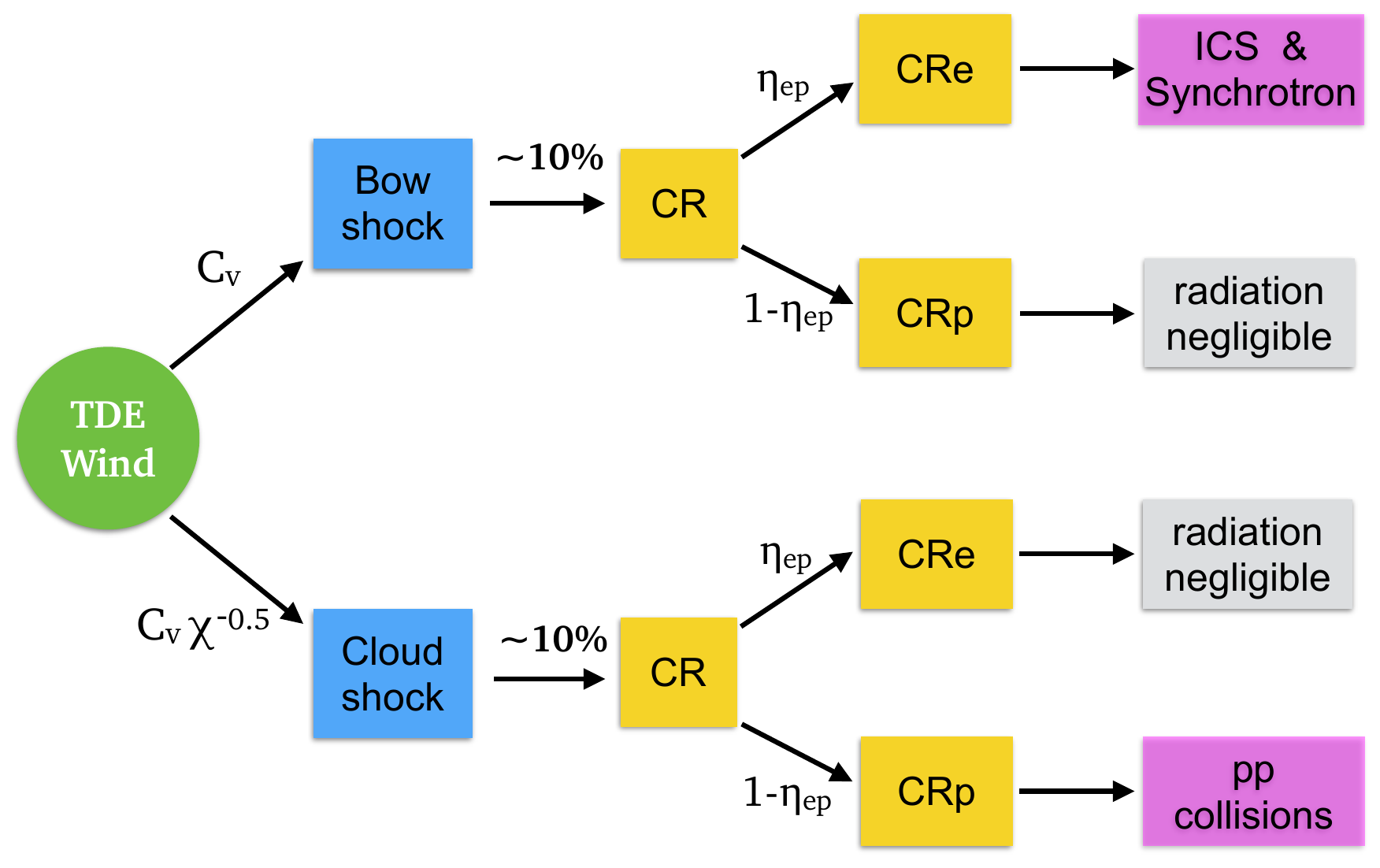}
 \caption{Schematic diagram of energy conversion relationship. When the TDE wind interacts with the AGN torus with a covering factor of $\cv$, the bow shock will take a fraction of $\cv$ of the wind's kinetic energy, while the cloud shock takes $\cv \chi^{-0.5}$ of the wind's kinetic energy. About $\sim 10\%$ of the shock energy is converted into CR's energy, among which 
 a fraction of $\eta_{ep}$ is taken by CRe, and
the rest is occupied by CRp. As analysis in the main text, the ICS of CRe at bow shocks dominates the gamma-ray radiations due to the strong AGN radiation field there (magenta box). The synchrotron of CRe at bow shocks dominates the 
{\bf radio--infrared } 
emission. Neutrinos are mainly contributed by pp collisions at cloud shocks due to the high gas density environment (magenta box). Radiations of CRp at bow shocks and CRe at cloud shocks are negligible (grey box). }   
 \label{fig3}
\end{figure*}

\section{TDE Wind and Torus Interactions}

Theoretical studies show that TDEs can drive strong wind in two possible processes.
One is self-interaction process when the general relativistic precession is strong (\citealt{sadowski2016, jiangyf2016}). 
The other one is the final settled accretion disk with high accretion rate. 
The kinetic luminosity can be $10^{44-45}~\ergs$, or even up to $10^{46}~\ergs$, while the mass outflow rate is up to a few solar mass per year (\citealt{dai2018, curd2019, lu2020}). 
The velocity of TDE wind is up to $10^4 ~\kms$. 
Such strong wind only lasts for months since both the strong general relativistic precession and the high accretion mode are short-termed.
 On one hand, 
according to a global simulation work on TDE of an IMBH ($10^5$ solar mass, \citealt{sadowski2016}), the duration of self-crossing-induced wind is comparable to (three times therein if more accurately) the orbital period of the most bound debris, which should be in the order of $\sim 1$ month if the result can be extrapolated to a SMBH.
On the other hand, the high accretion rate of the settled accretion disk is up to a few solar mass per year, and the total mass of the bound debris limit the high accretion rate to only maintain months. Thus, the duration of the TDE wind in this work is assumed to be 2 months, after which strong TDE wind is no long launched. Other fiducial parameters are listed in Table \ref{table1}.  

For a simplified wind ejected in spherical symmetry, the density follows 
$\rho_{\rm w}(r)=\dot M_{\rm w}/(4\pi r^2 \vw)=3.2\times 10^{4} ~\mhcm \times $ $(\dot M_{\rm w}/1 \msunyr) (\vw/10^4 \kms)^{-1} (r/0.1{\rm pc})^{-2}$. 
When the fast TDE wind encounters a dense cloud, a bow shock forms at the windward side of clouds, and vanishes as the transient wind disappears. 
In the meanwhile, the encounter of wind and cloud also drives a cloud shock inside the cloud with a velocity of 
$\vsc \simeq\chi^{-0.5} \vw$, where $\chi \equiv \rhoc/\rhow$ is the density contrast between the cloud and the wind (\citealt{mckee1975}; we caution that the expression in \citet{mckee1975} has different meaning, and details are presented in our appendix \ref{wdcld}). 
The velocity of the cloud shock is much lower than the wind velocity, but still can reach $\sim 1000~\kms$. The timescale of cloud shock sweeping across one cloud is $2\Rc/v_{\rm s,c} = 0.6 {\rm yr} (\Rc/10^{15}{\rm cm})(v_{\rm s,c}/1000\kms)^{-1}$ in which $2\Rc$ is the size of the cloud. 
The cloud size is quite uncertain in the present.  
Observations by X-ray eclipse events or water maser suggest that the size of the cloud may be around $10^{15}$cm (e.g., \citealt{kondratko2005, rivers2011}; \citealt{markowitz2014}). Geometrical models for fitting IR spectra adopt larger sizes of clouds (e.g., \citealt{honig2006, nenkova2008, stalevski2012}). 
Thus, after the TDE wind passes the cloud, the cloud shock continues to propagate inside the cloud for years
 (see appendix A.1. in \citealt{mou2021}).  
In the meanwhile, the radiative cooling timescale of post-shock cloud materials is 
$\tc \sim 1.8 {\rm yr} ~ T_{7} (\Lambda/10^{-23}{\rm erg ~cm^{3}~ s^{-1}})^{-1} n^{-1}_{7}$, in which $\Lambda$ is the cooling function (\citealt{sutherland1993}), $T_7\equiv T/10^7 K$ and $n_7 \equiv n/10^7\cmc$. The lifetime of cloud shock is limited by the minor one in above two timescales, and we argue that the cloud shock exists for about 
one year.  

In short, the bow shock is transient and exists for months, while its velocity equivalent to the wind velocity is very high (up to $10^{4} ~\kms$). The cloud shock is relatively more lasting and exists for $\sim$1 year. The velocity of the cloud shock is much lower, which may be around $1000 ~\kms$. 
Therefore, CRs at bow shocks and cloud shocks should be treated separately.

\section{Cosmic Rays}
\subsection{Acceleration of Cosmic Rays}

Shock can accelerate the charged particles to relativistic ones (cosmic rays) efficiently by the first order Fermi acceleration mechanism. 
The energy spectrum of cosmic ray follows a power law form of energy: $dN(E)/dE \propto E^{-\Gamma}$. The power law index is $\Gamma=1+3/(r-1)$, where $r$ is the compression ratio of downstream density to upstream density. According to the Rankine-Hugoniot condition, the compression ratio is $r=(\gamma+1)/(\gamma-1+2/M^2)$ where $M$ is the Mach number. For an adiabatic index of $\gamma=5/3$ and strong shocks ($M \gg 1$), the index is $\Gamma \approx 2.0$. 

Assuming that CR's diffusion in upstream/downstream is Bohm diffusion and ignoring the cooling process, the maximum energy of the particle is determined by (e.g., \citealt{reynolds2008}):
\be 
E_{\rm max}\approx 1 ~{\rm TeV} Z B_{\rm u,mG} v^2_{\rm s,3} t_{\rm acc, yr}
\ee
where $Z$ is the charge number of the particle, $B_{\rm u,mG}$ is the magnetic field in the upstream in mGauss, $v_{\rm s,3}$ is the shock velocity in units of $10^3~\kms$ and $t_{\rm acc,yr}$ is the acceleration time in units of year. 

The magnetic field in a cloud with a density of $10^{6-7}~\mhcm$ is typically $\sim 10^{0}$ mG (\citealt{crutcher2010}). 
Moreover, for the well studied Northern Arm structure at central sub-parsec in our galaxy, the magnetic field is $\sim 2$ mG (\citealt{roche2018}).  
Thus, we argue that $B_{\rm u, mG} \sim 10^{0}$ in the torus clouds. 
However, the magnetic field in upstream of the bow shock is quite unclear. 
Theoretically, the magnetic field strength can be amplified due to resonant streaming instability excited by relativistic particles (\citealt{bell2001, bell2004, schure2012}). 
The magnetic field amplification is also verified by observations of young supernova remnants,  
in which the magnetic pressure is $\sim 10^{-2} \rhow \vw^2$ in downstream (\citealt{volk2005}) and $\sim 10^{-3} \rhow \vw^2$ in upstream (e.g., \citealt{morlino2012}). 
Considering the ram pressure of the bow shock is $\rhow \vw^2=0.05 ~{\rm erg~ cm^{-3}}~(\dot M_{\rm w}/1 \msunyr) (\vw/10^4 \kms) (r/0.1{\rm pc})^{-2}$, we have $B_{\rm u} \sim 35 {\rm mG} (\dot M_{\rm w}/1 \msunyr)^{0.5} (\vw/10^4 \kms)^{0.5} (r/0.1{\rm pc})^{-1}$ in upstream of the bow shock, and $\Bd \sim 110{\rm mG} (\dot M_{\rm w}/1 \msunyr)^{0.5} (\vw/10^4 \kms)^{0.5} (r/0.1{\rm pc})^{-1}$ in downstream of the bow shock. 
Here, we adopt $B_{\rm u}=30$ mG and $\Bd=100$ mG  at bow shocks
as the fiducial parameters. 

If the acceleration timescale $t_{\rm acc}$ is comparable to the duration of the shocks, we would expect a maximum CR energy of several PeV at the bow shock and $\sim$TeV at the cloud shock. However, when coolings are included, $t_{\rm acc}$ is determined by the minimal value among the shock lifetime and cooling timescales.

\subsection{Energy Loss Processes for Cosmic Rays}
Ignoring the diffusion, the evolution of the energy distribution of CRs follows 
\be
\frac{\partial \Ncr(\Ecr, t)}{\partial t}-\frac{\partial}{\partial \Ecr}\left[ \dot{E}_{\rm cr} \Ncr(\Ecr, t) \right]= S(\Ecr,t)
\label{evolsed}
\ee
in which $\Ecr$ is the CR's energy, $\Ncr(\Ecr, t) d\Ecr$ represents the differential number of CRs between $\Ecr$ and $\Ecr+d\Ecr$ at time $t$, $\dot{E}_{\rm cr}$ is the total energy loss rate of CRs, $S(\Ecr,t)$ is the source function, and the subscript ``cr'' can represent ``crp'' or ``cre''. 
$\dot{E}_{\rm cr}$ of CRp is dominated by pp collisions and adiabatic loss: 
\be
\dot{E}_{\rm crp}=\dot{E}_{pp}+\dot{E}_{\rm ad},
\ee
in which $\dot{E}_{pp}$ is the pp collision cooling term and $\dot{E}_{\rm ad}$ is the adiabatic loss term. 
The timescale of pp collisions is (e.g., \citealt{aharonian2004})
\be
t_{pp}=(\nh \sigma_{pp} f c)^{-1} \simeq 5 ~{\rm yr}~ n^{-1}_7 , 
\ee 
 in which $\sigma_{pp}$ is the cross section of pp collisions, 
$f \simeq 0.5$ represent the coefficient of inelasticity, and $\nh$ is the number density of protons ($n_7\equiv \nh/10^7 \cmc$). 
 For $E_{\rm crp} \gtrsim 2$ GeV, the cross section can be approximated as $\sigma_{pp} \approx 30 ~{\rm mb} [0.95 + 0.06 \ln(E_{\rm crp}/1 {\rm GeV})]$ (\citealt{aharonian2004}) in the GeV to TeV energy region. 
At the bow shock, $t_{pp}$ is much longer than the existence timescale of bow shock, and $\dot{E}_{pp}$ is negligible, while at the cloud shock, $t_{pp}$ is slightly longer than the existence timescale of cloud shock. 
The adiabatic cooling will be discussed below as CRp and CRe share the same process. 

For CRe, the cooling is more complex and includes more processes:
\be
\dot{E}_{\rm cre}=\dot{E}_{\rm IC} + \dot{E}_{\rm syn} + \dot{E}_{\rm brem} + \dot{E}_{\rm ad},
\ee
in which the four terms in the right side represent the inverse Compton (IC) cooling, synchrotron cooling, bremsstrahlung cooling and adiabatic loss, respectively. 

First, the IC cooling timescale is given by 
 \be
  t_{\rm IC}= E_{\rm cre}/\dot{E}_{\rm IC} ,
\ee
where $\dot{E}_{\rm IC}= \int dN_{\gamma}(\Egm)/d\Egm \cdot \Egm d\Egm$, and $dN_{\gamma}(\Egm)/d\Egm$ is the number of photons produced per unit time per unit energy from ICS of CRe (see equation \ref{icsdnde} in appendix \ref{ics}). 
When the photon's energy before scattering in the electron rest frame is much less than $m_e c^2$ (Thomson limit), the expression can be simplified as $t_{\rm IC}=3\times 10^7 {\rm s}~\gamma^{-1} u^{-1}_{\rm ph}$,   
in which $\gamma$ is the Lorentz factor of CRe and $u_{\rm ph}$ is the radiation energy density in $\ergcm$. 
The distance from the central BH to the inner edge of the torus $r_{\rm t}$ is determined by the dust sublimation radius --- $0.1 {\rm pc} ~L_{\rm AGN,44}^{0.5}$ (\citealt{netzer2015}). Thus, if it is assumed that the AGN has been restored to its previous luminosity when the wind reaches the torus inner edge, we will have $u_{\rm ph} = L_{\rm AGN}/(4\pi r^2_{\rm t} c)=2.78\times 10^{-3}~\ergcm$, which is a constant. 
In particular, for CRe at cloud shocks, the post-shock cloud materials are almost fully ionized and can be treated as transparent for AGN radiations. As an approximation, we can still assume that the radiation field there is dominated by AGN radiation, and the IC cooling timescale shares the same form.  
 The IC cooling timescale is plotted in Figure \ref{fig2}, and it deviates from $t_{\rm IC}\propto E^{-1}_{cre}$ for $E_{\rm cre} \gtrsim 10^{11}$ eV due to the Klein-Nishina suppression.  

Second, the cooling timescale of synchrotron radiation in magnetic field B (in Gauss) is:
$t_{\rm syn}=8\times 10^8 {\rm s} ~ \gamma^{-1} B^{-2}$. 
The magnetic field in downstream is usually larger than the upstream since the magnetic field will be compressed together with the materials by the shock. \citet{caprioli2014} find that the downstream magnetic field can reach 4--10 times the undisturbed upstream magnetic field by simulations. Thus, the synchrotron cooling mainly occurs in downstream of the shock. 
Taking $\Bd \simeq 10^{2}$ mG at bow shocks, and 10 mG at cloud shocks, the synchrotron radiation timescale of CRe is 
\be
t_{\rm syn}= 4\times 10^{7}{\rm s} \left(\frac{E_{\rm cre}}{1\GeV}\right)^{-1} \left(\frac{\Bd}{100{\rm mG}}\right)^{-2} 
=  4\times 10^{9}{\rm s}\left(\frac{E_{\rm cre}}{1\GeV}\right)^{-1} \left(\frac{\Bd}{10{\rm mG}}\right)^{-2}. 
\ee

Third, for the bremsstrahlung of CRe, the cross section is $\sigma_{\rm br} \sim 20-30$mb (appendix \ref{bremcre}), and the cooling timescale is 
\be
t_{\rm br}=(\nh \sigma_{\rm br} c)^{-1} \simeq 5~{\rm yr} ~ n^{-1}_{7} .
\ee
At bow shocks, due to the low density environment of $n_{7}\ll 1$, $t_{\rm br} \gg t_{\rm IC}$. 
At cloud shocks, this timescale can be comparable to $t_{\rm IC}$ of GeV electrons only if the cloud density is as high as $\ga 10^{9}\mhcm$, which is a harsh condition for torus cloud. Thus, bremsstrahlung of CRe can be ignored.  

Fourth, for the adiabatic process of relativistic particles, the energy loss rate as they do work in expanding is (\citealt{longair1994}): 
$\frac{d\Ecr}{dt}= -\frac{1}{3} (\nabla \cdot \vec{v}) \Ecr $. 
At the bow shock, we write the adiabatic cooling timescale as:
\be
t_{\rm ad}=\Ecr/|d\Ecr/dt| = \kbow \Rc/\vw = 1\times 10^6 {\rm s} ~\kbow v^{-1}_{\rm w, 4} R_{\rm c,15}
\ee
in which $\Rc/\vw$ is the dynamical timescale, and $\kbow$ is a dimensionless coefficient. We argue that the motion of CR in post-shock stream cannot be treated as a free expansion, but instead should be strongly confined by shocked wind. Simulation tests on wind-cloud interactions suggest that $\kbow \sim 10$ (appendix \ref{adb}, when the duration of the wind is $\gg \Rc/\vw$, $\kbow$ approaches 20), indicating that the adiabatic cooling timescale is much longer than the dynamical timescale. For $R_{\rm c,15}= 2.5$ and $\vw = 3\times 10^4~\kms$, we have $t_{\rm ad}\simeq 3$ months. Thus, if we only concern the radiations within the first few months of wind-torus interactions in this work, ignoring the adiabatic loss will not bring too much error.   
Inside the cloud, the adiabatic loss is 
 moderate and is around one year (see appendix \ref{adbcld} for simulation tests). 
Thus, adiabatic cooling of CRs at cloud shock can also be ignored within the first few months of wind-torus interactions. 

The acceleration timescales and various energy loss timescales as functions of CR energy are plotted in Figure \ref{fig2}, from which we can obtain the maximum energy of CRp and CRp at both shocks.  
The maximum CR energies at bow shocks are $(\Emaxp, \Emaxe) = (10^3~\TeV, 1~\TeV)$ for the fiducial parameters, 
while at cloud shocks, we have $(\Emaxp, \Emaxe) = (1~\TeV, 10~\GeV)$.

\begin{table*}
  \centering
%  \begin{minipage}{120mm}
  \renewcommand{\thefootnote}{\thempfootnote}
  \caption{Fiducial Parameters in Our Model.  }
  \begin{tabular}{@{}  c  c  c  c}
  \hline
  \hline
         Parameters
        & Descriptions
        & Fiducial Values  \\
    \hline
      $\cv$      & covering factor of the torus  &  0.4  \\ % 
     $\chi$   & density contrast $\rhoc/\rhow$ &  300   \\ %
     $\rhoc$ ; ~ $n_7$    & cloud density; ~ $n_7\equiv \rhoc/10^7 m_{\rm H}$  & $10^7 \mhcm$   \\
     $\Rc$; ~$R_{\rm c,15}$     & cloud radius; ~ ~$R_{\rm c,15}\equiv \Rc/10^{15}$cm & $2.5\times 10^{15}$cm \\
    $\Lkin$ ; ~ $\Lkinf$ & kinetic luminosity of TDE wind;  $\Lkinf\equiv \Lkin/10^{45}\ergs$ &  $1\times 10^{45} ~\ergs$ \\
    $\dot M_{\rm w}$ & mass outflow rate of TDE wind & $3.5\msunyr$ \\
    $\vw$    & velocity of TDE wind & $3\times10^4\kms$ \\
    $\rhow$  & density of TDE wind at the inner edge of the torus & -- \\
    $\Lagn$ & bolometric luminosity of AGN & -- \\ 
    $t_{\rm bst}$ & duration of TDE wind & 2 months \\
    $\Kep$   &  $K_{\rm inj,e}/K_{\rm inj,e}$ & 0.01 \\
   $\ecre$  & total energy of CRe without cooling (equation \ref{ecr})  & -- \\
   $\ecrp$  & total energy of CRp without cooling (equation \ref{ecr})  & -- \\
   $\vsc$ & cloud shock speed & 1000 $\kms$  \\ 
   $\Bd$  & magnetic field strength in downstream of bow shocks & 100 mG \\
    \hline
 \label{table1}
 \end{tabular}
\end{table*}

\subsection{Spectral Energy Distribution of CRs}

As shown in Figure \ref{fig3}, we only consider following major radiations: 
ICS and synchrotron radiation of CRe at bow shocks and pp collisions of CRp at cloud shocks, while ignoring radiations of CRp at bow shocks (due to the low gas density there) and radiations of CRe at cloud shocks (due to the low total energy of CRe there). Quantitative analysis is presented in Section 4.  

Assuming that during the shock acceleration stage, there is continuous injection of CRp and CRe with an energy spectrum of $S(E,t) = \Kinj E^{-\Gm}$ ($\Emax \leqslant$ 1 TeV, $\Gm=2$). Here we do not include the dependence of the maximum energy on the injection time.
For comparison, we also test softer spectral indices of $\Gamma=2.5$ and $3.0$ (\citealt{chevalier2006}). 

For CRp (cloud shock), the injection can be roughly treated as a stationary process without cooling if the time concerned does not exceed one year. The spectral energy distribution of CRp at time $t$ follows:  %($\leqslant$ the life of cloud shock)
\be
N_p (E_p, t)=K_p(t) E^{-\Gamma}_p \exp(-E_p/\Emax) ~~, 
\label{eqNp}
\ee
in which $K_p(t)=K_{\rm inj,p} t$. 

For CRe
 (bow shock),  
however, the spectral index will change with time since the cooling timescale of CRe with energy $\ga 10^0$ GeV is shorter than the lifetime of bow shock. 
As mentioned above, the energy loss rate of CRe is dominated by ICS in our fiducial model. The ICS cooling rate can be fitted by 
$\dot{E}_{\rm cre}=\left[ (a E_e^2)^{-1}+(b E_e^{\delta})^{-1}+c \right]^{-1}$ GeV s$^{-1}$,
where $E_e$ is in GeV, and for $E_e \leqslant 1$ TeV, the coefficients are: 
$a=1.66\times 10^{-7}$ GeV s$^{-1}$, $b=1.95\times 10^{-6}$ GeV s$^{-1}$, $\delta=1.4$, $c=0$ (the accuracy of this expression is no less than 93\% for $E_e \leqslant 1$ TeV). \footnote{We caution that, this expression only applies to the photon field at the inner edge of the dusty torus, and the shape of AGN spectrum is the composite one for an Eddington ratio of $10^{-3}<L_{\rm AGN}/L_{\rm Edd} < 10^{-1}$ (\citealt{ho2008}).}

1), in the case of $\dot E_e t \geqslant E_e$, namely, $E_e \gtrsim 2.3~\GeV (t/{\rm month})^{-1}$ \footnote{More accurately, $E_e \geqslant (2.3+0.2E^{0.6}_e) (t/{\rm month})^{-1}$, if we write $E_e$ in units of GeV. }, the source term is balanced by energy loss term, and we obtain a stationary distribution of CRe:
\be
~~~~~~ N_e(E_e, t) \simeq \frac{K_{\rm inj,e} E_e^{-(\Gamma+1)} (a^{-1}+b^{-1}E^{0.6}_e)}{(\Gamma-1)}  ~.
\label{eqNe1}
\ee  

2), when $\dot E_e t < E_e$, namely, $E_e \lesssim 2.3~\GeV (t/{\rm month})^{-1}$, ICS cooling of CRe can be simplified to the Thomson limit if $t \gtrsim 0.5$ month. Thus, the distribution is 
\be
~~~~~~ N_e(E_e, t) \simeq \frac{K_{\rm inj,e} E_e^{-(\Gamma+1)}}{a(\Gamma-1)} \left[ 1-(1-a E_e t)^{\Gamma-1} \right]  ~.
\label{eqNe2}
\ee

The total energy of CR gained in shock acceleration without cooling is: 

$\epsilon_{\rm cr}= \int^{\infty}_{m_0 c^2} (\Ecr-m_0 c^2) \Ncr(\Ecr) d\Ecr$
\be
=
\left\{
\begin{split} 
 K_{\rm inj,p} t \int^{\infty}_{m_p c^2}  (E_p-m_p c^2) E_p^{-\Gamma} \exp(-E_p/E_{\rm max, p}) dE_p ~~~({\rm CRp}) \\
 K_{\rm inj,e} t \int^{E_{\rm max,e}}_{m_e c^2}  (E_e-m_e c^2) E_e^{-\Gamma} dE_e  ~~~~~~~~~~~~~~~~~~~~~~~~~~~~ ({\rm CRe})
\end{split}
\right.
\label{ecr}
\ee 
From the studies of SNRs (e.g., \citealt{hinton2009, blasi2013}) and numerical simulations (\citealt{caprioli2014}), about $10\%$ of the shock energy can be converted into CRs.
People use 
$\Kep \equiv K_{\rm inj,e}/K_{\rm inj,p}$ 
to describe the number ratio of CRe and CRp at a given energy or momentum. 
Ignoring radiative cooling of CRs, one dimensional simulations on DSA of protons and electrons with high Mach number, quasi-parallel, collisionless shocks show that $\Kep=10^{-2}- 10^{-3}$ (\citealt{park2015}), while 3-D simulations with quasi-perpendicular strong shocks report a much higher ratio where up to $\sim 10^{-2}$ of the \emph{shock-dissipated energy} is converted into CRe (\citealt{winner2020}). 
Observations of Tycho's SNRs suggest that $\Kep$ is $1.6\times 10^{-3}$ (\citealt{morlino2012}). 
Direct measurement around the Earth by PAMELA reports that $\Kep \simeq 0.01$ at 10 GeV (\citealt{picozza2013}). 
Observations of radio relics in galaxy cluster mergers suggest higher values: $\Kep=10-10^{-2}$ (\citealt{vazza2015}) or $\Kep \ga 0.1$ (\citealt{brunetti2014}). 
In this work, we adopt $\Kep=0.01$ as the fiducial parameter.

%%% =========================================
%%% =========================================

\section{Results}
We calculate the emissions of CRe/CRp accelerated at bow shocks and cloud shocks separately.
Details of radiation mechanisms are presented in appendix, including pp collisions (appendix \ref{ppc}), ICS of CRe (appendix \ref{ics}) and synchrotron of CRe (appendix \ref{syn}). By estimations, we find that other mechanisms such as bremsstrahlung of CRe (appendix \ref{bremcre}), ICS of secondary CRe originated from pp collisions (appendix \ref{2ndcre}), and photomeson production in p$\gamma$ reactions (appendix \ref{pgm}) are very weak and can be ignored.  

Moreover, the photon-photon pair production ($\gamma \gamma \rightarrow e^{-} e^{+}$) is an important process for gamma-ray absorptions. The threshold energy is given by $\Egm \geqslant 2 m^2_e c^4/[h\nu (1-\cos\alpha) ]$, where $h\nu $ is the target photon energy and $\alpha$ is the collision angle. Thus, gamma-rays of $\ga 10^2$ GeV are able to interact with optical/UV photons from AGN. 
The cross-section for pair production has a maximum at the level of $0.2\sigma_T$ (e.g., \citealt{aharonian2004}). 
Due to the high radiation energy density at the torus inner edge, the AGN photons there are quite abundant, of which the optical/UV photon density is around $10^8 ~\cmc$. Considering the size of the pair production region of $\sim r_t\sim 10^{17}$ cm, gamma-rays of $\Egm \ga 10^2$ GeV may be strongly absorbed. Thus, we include the photon-photon interactions in calculating gamma-ray spectra, and details are presented in appendix \ref{gmgm}. 

\subsection{Bow Shock}

As mentioned above, TDE wind is expected to be energetic with a kinetic luminosity of $\Lkin=10^{44-46}~\ergs$. Taking $\Lkin=10^{45}~\ergs$ for the fiducial value, when the wind is interacting with clouds with covering factor of $\cv$, the energy converted into CR (CRp+CRe) per unit time is $\sim 10\% \cv \Lkin =  1\times 10^{44} ~\ergs ~\cv \Lkinf$, in which $\Lkinf \equiv \Lkin/10^{45}\ergs$.
The energy injection rate for CRe is $\dot \epsilon_{\rm cre}=1\times 10^{42} ~\ergs (\Eep \cv/0.01) \Lkinf $. 
Ignoring the cooling, the total injected energy of CRe by time $t$ is 
\be
\ecre=\dot \epsilon_{\rm cre} t = 2.6~\times 10^{48} {\rm erg} ~ \Lkinf \frac{\Eep \cv}{0.01} \frac{t}{1 \rm month} ~. 
\label{bscre}
\ee 
In order to set the AGN radiation field, we use the composite AGN SED for an Eddington ratio of $10^{-3}<L_{\rm AGN}/L_{\rm Edd} < 10^{-1}$ (\citealt{ho2008}), and scale it to a bolometric luminosity of $10^{44}~\ergs$ (the luminosity value does not affect ICS of CRe, since the radiation energy density at inner edge of the torus is a constant due to $r_t \propto L^{0.5}_{\rm AGN}$). 
Taking $\Eep \cv =0.01$ as the fiducial value, we can calculate the gamma-ray spectrum contributed by ICS of AGN photons by CRe (Figure \ref{fig4sed}). The gamma-ray luminosity from ICS is up to $\sim 3\times 10^{41}~\ergs \Lkinf$ 
 for $\Egm > 0.1$ GeV,  
and is not sensitive to the time due to the ICS cooling saturates the high energy CRe of $> 10^0$ GeV. Moreover, the gamma-ray spectrum shows an obvious cut-off at tens of GeV due to the absorptions by AGN photon field via the photon-photon pair production. 
 For softer power-law indices of CRe, the expected gamma-ray luminosities will be lower, which are $1.1\times 10^{41}~\ergs$ and $7\times 10^{39}~\ergs$ for $\Gamma=2.5$ and 3.0, respectively (for $\Egm > 0.1$ GeV, see the lower panel in Figure \ref{fig4sed}). 

At bow shocks, assuming that the magnetic field is $\Bd$=100 mG, we obtain the synchrotron emission of CRe (Figure \ref{fig5syn}). The radio luminosity $\nu L_{\nu} $ at 1-10 GHz is $10^{38-39}~\ergs \Lkinf$. We also consider a stronger magnetic field of $\Bd=1000$ mG (here the cooling is dominated by synchrotron emission instead of ICS), 
and find that the radio luminosity at 1-10 GHz can reach $10^{40}~\ergs$. 
 Moreover, the synchrotron emission also extends to infrared, which can reach $\sim 10^{40}~\ergs$. This infrared emission is intrinsically different from the infrared echo. 

For CRp, the total energy is
\be
\ecrp=2.6~\times 10^{49} {\rm erg} ~ \frac{(1-\Eep) \cv}{0.1} \frac{\Lkin}{10^{45}\ergs} \frac{t}{\rm month}
\ee
Considering the density at bow shock of the order of $10^{4-5}~\mhcm$, the gamma-rays from neutral pions in pp collisions is 
$\sim f_{pp} \ecrp(t)/\tau_{pp} \sim 3\times 10^{38}~\ergs~ n_5 \Lkinf (t/{\rm month})$, where $f_{pp}$ is the energy fraction of primary CRp carried by gamma-rays which is $\sim 1/6$ (\citealt{hinton2009}).    
This is too weak compared to the pp reactions at the cloud shock (to be discussed in the next section), and thus we neglected the emissions from pp collisions at the bow shock (Figure \ref{fig3}). 

\subsection{Cloud Shock}
\begin{figure}
   \includegraphics[width=0.99\columnwidth]{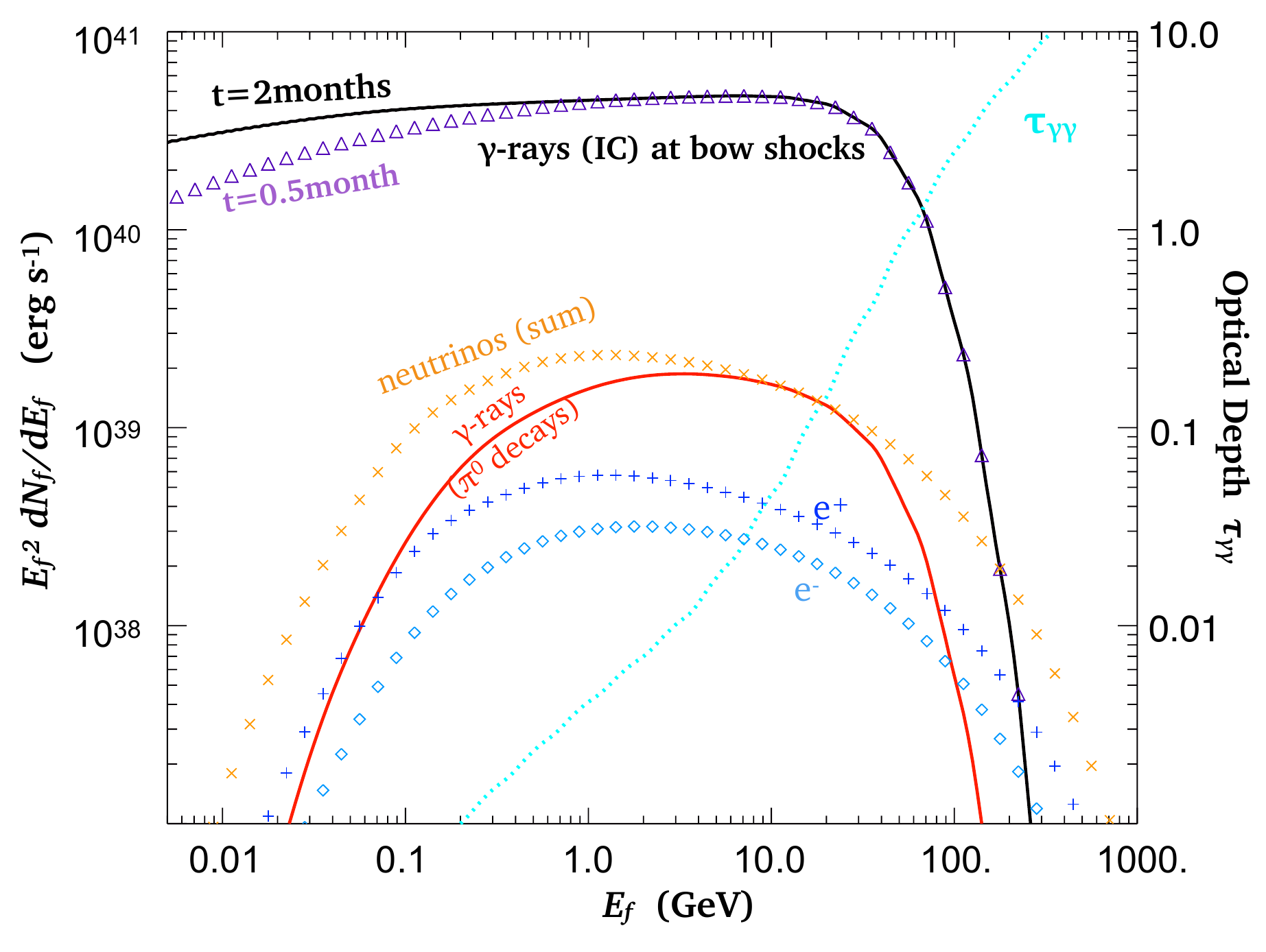}
   \includegraphics[width=0.93\columnwidth]{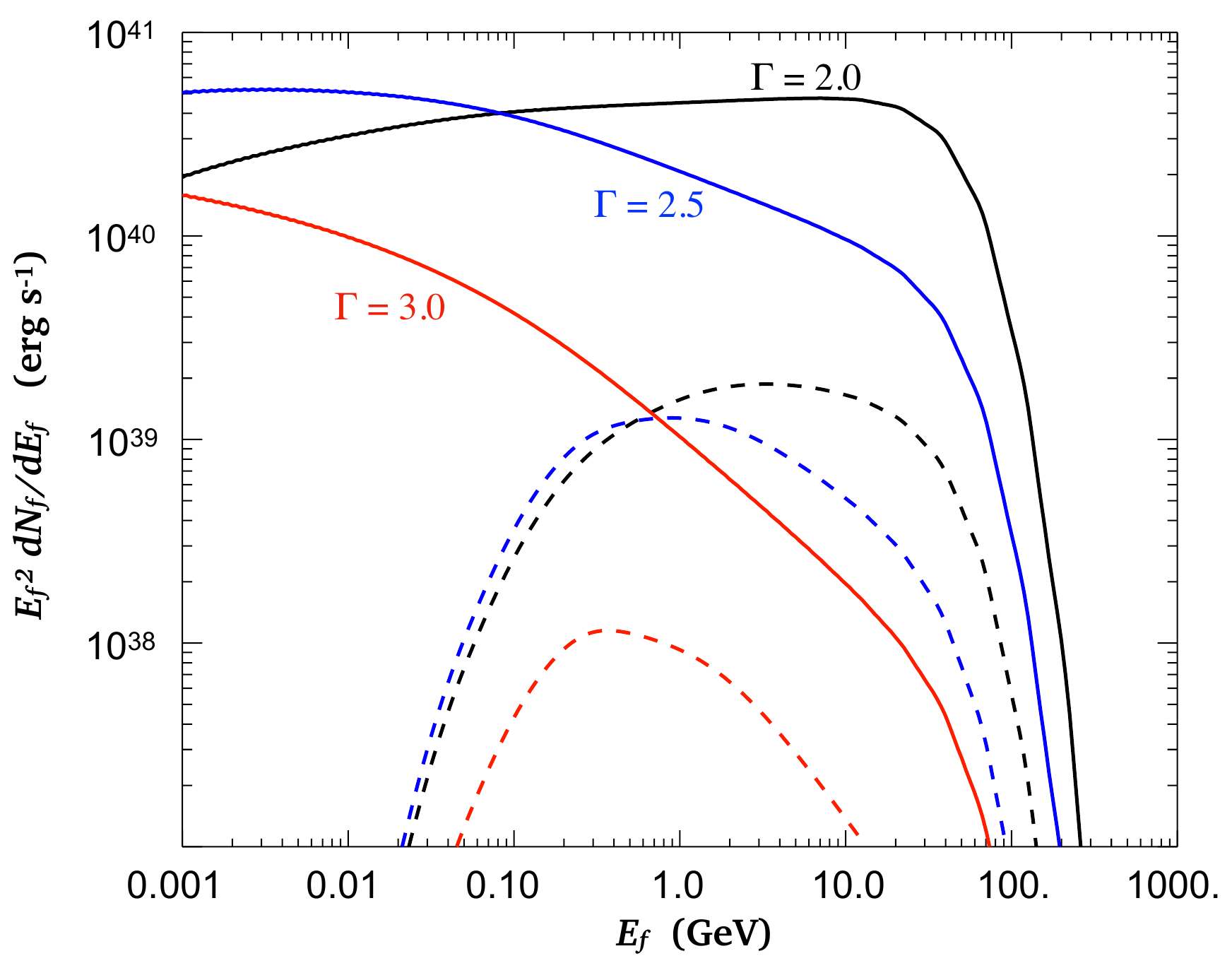}
   \caption{
 Upper panel: 
the spectral energy distributions of gamma-ray, secondary e$^{\pm}$ and neutrino (sum of $\nue, \anue, \num, \anum$) for the fiducial parameters. 
The gamma-rays are from ICS of CRe at bow shocks (black solid lines) and $\pi^0$ decays at cloud shocks (red solid lines). The solid lines mark the result including absorptions by photon-photon pair production 
 (optical depth $\tau_{\gamma\gamma}$ plotted in cyan dotted line). 
The gamma-rays are obviously dominated by ICS, of which $\Lgm = 2.8\times10^{41}~\ergs$ for $\Egm > 0.1$ GeV, while pp collisions at cloud shocks contribute 
 a luminosity of $\Lgm = 8.0 \times 10^{39} ~\ergs$. 
The neutrino emission from pp collisions is about $1\times 10^{40}\ergs$. 
For comparison, the gamma-ray spectrum from ICS at bow shocks at $t=0.5$ month is plotted with purple triangles. Due to the efficient IC cooling of high energy electrons, the gamma-ray emissions soon saturate in $\Egm \gtrsim 0.1$ GeV. 
 The lower panel shows the gamma-ray spectra for different values of $\Gamma$: 2.0 (black), 2.5 (blue) and 3.0 (red). Gamma-rays from ICS and pp collisions are plotted with solid lines and dashed lines, respectively. 
} 
 \label{fig4sed} 
\end{figure}

\begin{figure} 
\includegraphics[width=0.95\columnwidth]{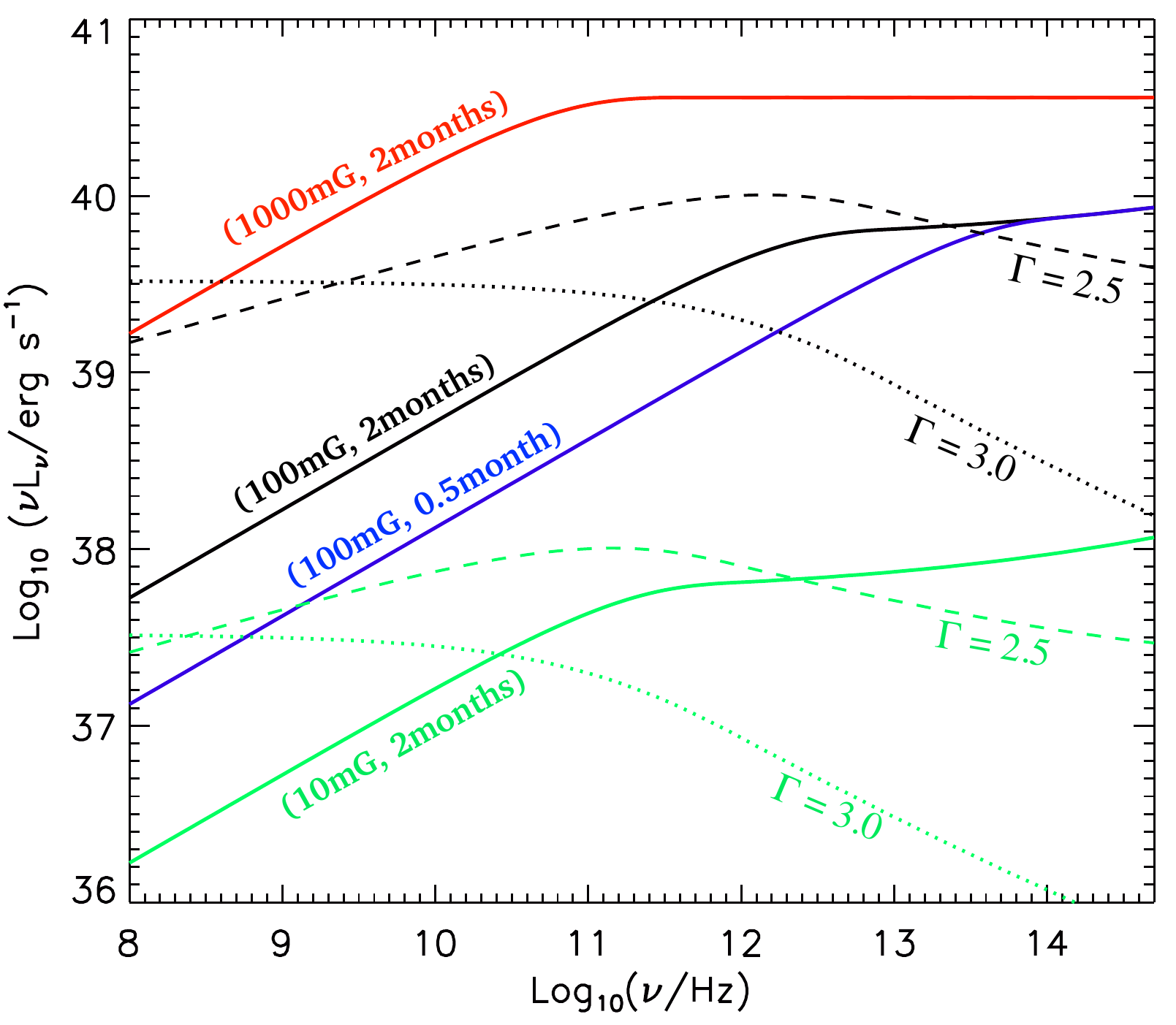}
 \caption{Synchrotron emissions  
 at bow shocks. The colors denote the values of $(B, t)$. The solid, dashed and dotted lines denote the values of $\Gamma=2.0, ~2.5$ and $3.0$, respectively. For the fiducial parameters (black solid line), the radio luminosity at $1-10$ GHz is $10^{38}~\ergs$, while it can reach $10^{39-40}~\ergs$ for softer power-law indices or stronger magnetic fields. Moreover, the synchrotron emission also contributes to infrared radiation, which can reach $\sim 10^{39-40}~\ergs$. 
} 
 \label{fig5syn}
% \vspace{0.2cm}
\end{figure}

Cloud shocks last for a longer timescale, which is typically $\sim$1 year. It's more convenient to analyze from the aspect of total energy. 
For large density contrast case ($\chi \gg 1$), only a small fraction of the wind's kinetic energy can be converted into the cloud's energy. 
The total energy driven into the cloud by wind is $E_{\rm c,tot} \simeq \cv \chi^{-0.5} \Lkin t_{\rm bst}$ (appendix \ref{wdcld}, see also \citealt{mou2021}).
The total energy of CR (CRp+CRe) at cloud shock is 
\be
E_{\rm CR}\simeq 0.1 E_{\rm c,tot} = 2.6\times 10^{48} {\rm erg}~ \frac{\cv \chi^{-0.5}}{0.01} \Lkinf \frac{t_{\rm bst}}{\rm month}.
\ee 
The energy of CRe is 
\be
\ecre = \Eep E_{\rm CR}=2.6\times 10^{47} {\rm erg} ~ \frac{\Eep \cv \chi^{-0.5}}{1\times 10^{-3}}  \frac{t_{\rm bst}}{\rm month} .  
\ee
This value is significantly lower than the bow shock (equation \ref{bscre}), and can be neglected. 
The energy of CRp is: 
\be
\ecrp =(1-\Eep) E_{\rm CR}=2.6\times 10^{48} {\rm erg} ~ \frac{(1-\Eep)\cv \chi^{-0.5}}{0.01} \Lkinf \frac{t_{\rm bst}}{\rm month}. 
\ee
Thus, the pp collision will give rise to a gamma-ray luminosity of
\be
\Lgm \sim f_{pp} \ecrp /\tau_{pp} \sim 3\times 10^{39}~\ergs \frac{(1-\Eep) \cv \chi^{-0.5}}{0.01}  n_{7} \Lkinf \frac{t_{\rm bst}}{\rm month}  , 
\ee
which is much lower than the luminosity from ICS of CRe at bow shocks. 
If the cloud density is $n_7 \gg 1$, the gamma-ray luminosity from pp collisions may be comparable to or even higher than the ICS of CRe at bow shocks.  
Taking the fiducial parameters, we obtain the SED of secondary particles including gamma-rays, electrons/positrons, neutrinos from pp collisions, which is plotted in 
 the upper panel in 
Figure \ref{fig4sed}. Consistent with the above analysis, the gamma-ray luminosity from pp collisions is $\sim 1\times 10^{40}~\ergs$, and the neutrino power is also $\sim 1\times 10^{40}~\ergs$. 

%%%%%%%%%%%%%%%%%%%%%%%%%%%%%%%%%%%
%%%%%%%%%%%%%%%%%%%%%%%%%%%%%%%%%%%

\section{Conclusions and Discussions}

In the scenario of strong TDE wind-torus interactions, we find that the dramatic collision is able to accelerate charged particles to relativistic energies, and generate gamma-ray, 
radio-infrared emissions and neutrinos. 
These emissions are multi-band and multimessenger afterglows accompanied with the X-ray afterglows with a time lag of $r_t/\vw \simeq 3 {\rm yr} ~\frac{r_t}{0.1 {\rm pc}} \left(\frac{\vw}{0.1 c}\right)^{-1}$, where $r_t$ can be measured by the time delay of infrared echo, the line width of the transient coronal lines or other methods.   

The gamma-rays are dominated by ICS at bow shocks, and the luminosity is 
$\sim 3\times 10^{41} ~\ergs \Lkinf$. For non-jetted AGN, gamma-rays have not been detected, while for blazars, the gamma-ray luminosity can reach $\sim 10^{44-45}~\ergs$ in 0.1-100 GeV (\citealt{padovani2017}). 
For star formation galaxies, the detected gamma-ray luminosities are $10^{37}-10^{42}~\ergs$(\citealt{ackermann2012}). 
Thus, the gamma-rays from strong TDE wind - torus interactions may be outstanding compared with those normal/active galaxies except blazers. 
In our fiducial model, the gamma-ray generation rate of $\Egm \geqslant 1$ GeV is $3.1 \times 10^{43}$ photons s$^{-1}$. If the source is located at a distance of 10 Mpc, the gamma-ray flux is $2.6 \times 10^{-9} ~{\rm photons ~s^{-1} ~cm^{-2}}$.  
For Fermi-LAT, the effective area in the center of the field of view is $\sim$7000 cm$^2$. In standard sky-survey mode, a given point on the sky only is viewed for $\sim$30 minutes every 3 hours (\citealt{atwood2009}). Thus, for a gamma-ray afterglow lasting for two months, the number of source gamma-rays ($\Egm \geqslant 1$ GeV) collected by Fermi-LAT is expected to be 16. 
The background gamma-ray flux above 0.8 GeV for $|b|\geqslant 10^{\circ}$ is $\sim 3 \times 10^{-6}~{\rm photons ~s^{-1} ~cm^{-2}~ sr^{-1}}$ (\citealt{abdo2010}). 
Taking the on-axis angular resolution of Fermi-LAT as $\sim 1^{\circ}$ above 1 GeV (\citealt{atwood2009}), the background photons within the angular resolution during the same time is $\sim$17. Thus, the signal-to-noice is only 16/17. Considering the effects of broad-line region and geometrical light travel to be mentioned below, we argue that the gamma-ray afterglows around 10 Mpc can not be detected by Fermi-LAT. The volume TDE rate of $10^{-8}-10^{-5}$ Mpc$^{-3}$ yr$^{-1}$ (e.g., \citealt{wang2004, van2014, stone2016}) puts an upper limit of 1 case per century for the detection rate by Fermi-LAT. 

The radio emission from the synchrotron radiation is also mainly contributed by bow shocks, and at 1-10 GHz, the luminosity is $10^{38-39}~\ergs \Lkinf$ for a magnetic field strength of 100 mG. 
 We note that some TDE candidates exhibit radio emission of $10^{36-42}~\ergs$ with time lags spanning from days to years (see review by \citealt{alexander2020}).  
The CRe for generating those radio emissions are thought to be accelerated in forward/external shocks driven by wind in the diffuse ISM, or reverse/internal shocks driven by jets in previous models. 
According to our results, we propose another scenario, in which for some radio flares lagging the primary outburst by months or years, their radio emissions may originate from the process of TDE wind impacting the clouds. Besides, the synchrotron emission also extends to infrared with a luminosity of $10^{39-40}~\ergs$. 

The GeV neutrino emission is $1\times 10^{40}~\ergs$, indicating that the flux is $<10^{-12}~\ergs$ cm$^{-2}$ when the source is beyond 10 Mpc. Such a neutrino flux is far below the sensitivity of the IceCube, and this is not conflict with the non-detection of transient neutrino emissions in 1-100 GeV band (\citealt{abbasi2020}). 
Moreover, due to little contribution to gamma-ray and neutrino luminosities, in this work we did not pay attention to the CRp at bow shocks. 
However, the maximum energy of those CRp can be accelerated to several PeV or even higher if the magnetic field is stronger. Thus, we would expect that neutrinos of $\sim 10^{2}$ TeV could be produced, although the quantity should be very low.  

We did not include the adiabatic cooling in obtaining above results. However, in the fiducial case, the adiabatic cooling timescale can be as short as 3 months for CRe at bow shocks. By the time $t=2$ months when the wind-cloud interactions just cease, CR's energy at bow shocks has been lost by one half during adiabatic process (see Figure \ref{figB2}). 
Thus, the above radiations from CRe at the bow shock may only last for a few months, after which they would be significantly weakened due to adiabatic cooling. The adiabatic cooling itself does not affect the spectral energy index of CRs, but translates the CR's SED down along the vertical axis ($N_{\rm cr}$) as a whole. Therefore, while the shapes of gamma-ray and radio spectra remain unchanged, the spectra will shift downward. 
The adiabatic cooling limits the duration of the gamma-ray and 
 radio--infrared
afterglows in the order of months (the specific timescale depends on $\kbow \Rc/\vw$), which may be significantly shorter than the duration of X-ray afterglows (\citealt{mou2021}).  

\begin{figure} 
\includegraphics[width=1.\columnwidth]{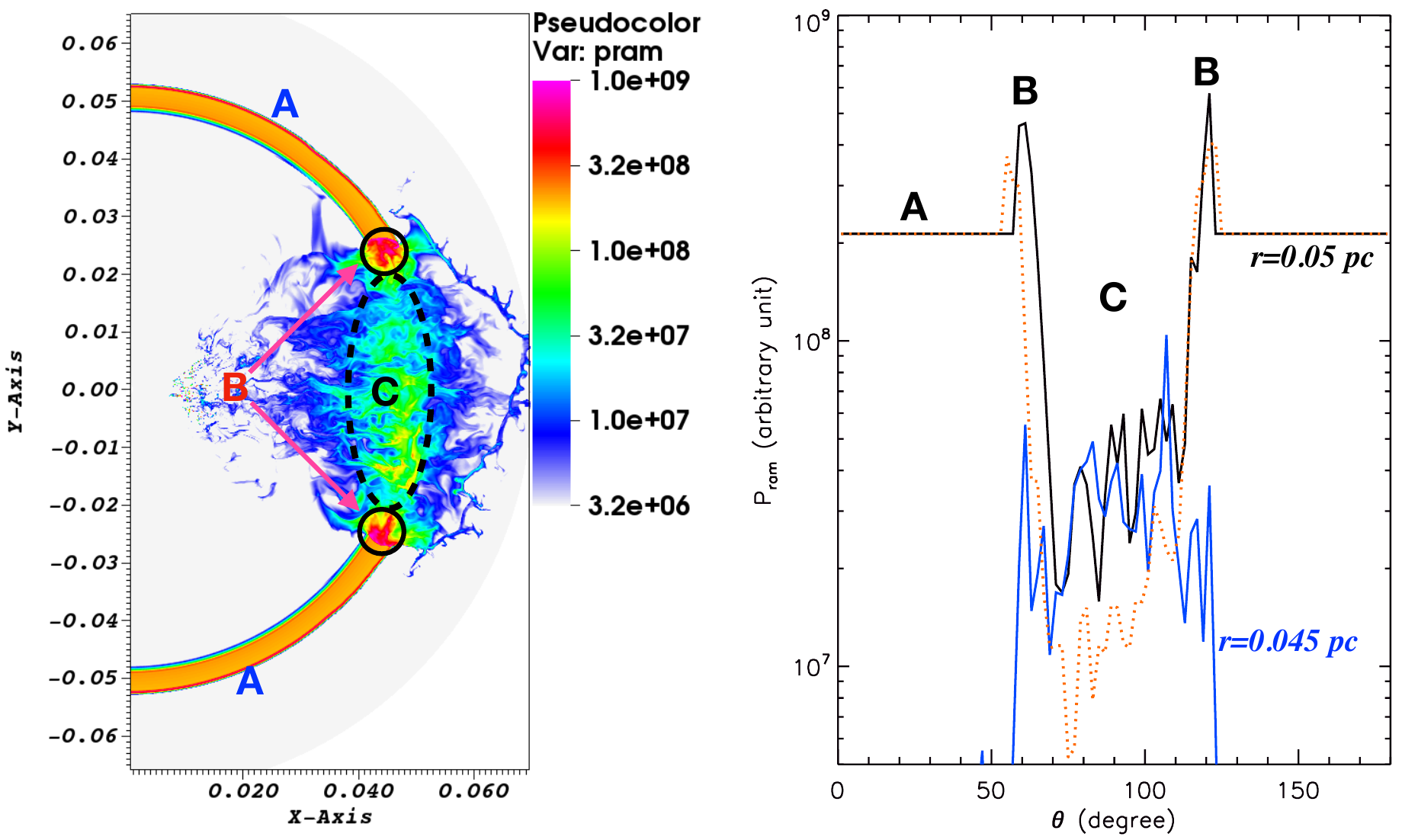}  % pub3/afg_blr/b110Hd59
 \caption{Effect of wind-BLR interactions. Left panel: distribution of the ram pressure in arbitrary units. 
After passing the BLR, the TDE wind is strongly disturbed, and forms a weakened region (region C) and an enhanced region (region B).  Right panel: distribution of $P_{\rm ram}$ along $\theta-$direction at different distances (0.050 pc and 0.045 pc, smoothed by $\Delta\theta=2^{\circ}$). In addition, the orange dotted line denotes the result of $\rho_{\rm BLR}=3\times 10^{10} ~\mhcm$ at $r=0.05$ pc. }
 \label{fig6blr}
\end{figure}

Before reaching the dusty torus, the TDE wind will first collide with the broad-line region (BLR), which is located closer to the central black hole. Intuitively, such an interaction will reduce the strength of the TDE wind impacting the torus. 
In order to explore this quantitively, we performed hydrodynamic simulations as a test, rather than a thorough study. In order to embody the BLR, we simply assume that the BLR is composed of a lot of small clouds with a density of $5\times 10^9~\mhcm$ (see appendix \ref{blr} for more details). After passing the BLR, the initially isotropic TDE wind is divided into three regions (Figure \ref{fig6blr}):
region A --- an undisturbed region; 
region B --- an enhanced region where the TDE wind is gathered after being blocked by the BLR;   
region C --- a weakened region where the TDE wind directly blocked by the BLR.  
In region B, the ram pressure of the TDE wind (equivalent to the volume density of its kinetic energy) is higher than that of region A, and its averaged value is $P_{\rm ram, B} \approx 1.7 P_{\rm ram, A}$. % (averaged within $r=0.049-0.053$pc). 
In region C, the ram pressure is significantly weakened, and the averaged value is $P_{\rm ram, C}\approx 0.2 P_{\rm ram, A}$. 
The global averaged ram pressure in both region B and C is about 50\% of region A, inducing that after being blocked by BLR, the averaged strength of the TDE wind is significantly weakened. The expected radio and gamma-ray luminosity would correspondingly drop by $\sim 50\%$. 
If the BLR's density is increased to $3\times 10^{10}~\mhcm$, the averaged ram pressure of region B+C would decrease to 29\% of region A. If region B just misses colliding with the torus, the expected gamma-ray and radio luminosity would drop by an order of magnitude. Thus, if there a BLR, the TDE wind-torus interactions would become weaker, which depend on the mass of BLR clouds, and the spatial distributions of BLR and torus clouds. 
Besides, the wind-BLR interactions would lead to radiations with shorter delays. However, the strength is very dependent on the physics of BLR clouds, which is quite uncertain. Here we did not explore it further (see appendix \ref{blr} for further discussions). 

\begin{figure} 
\includegraphics[width=0.95\columnwidth]{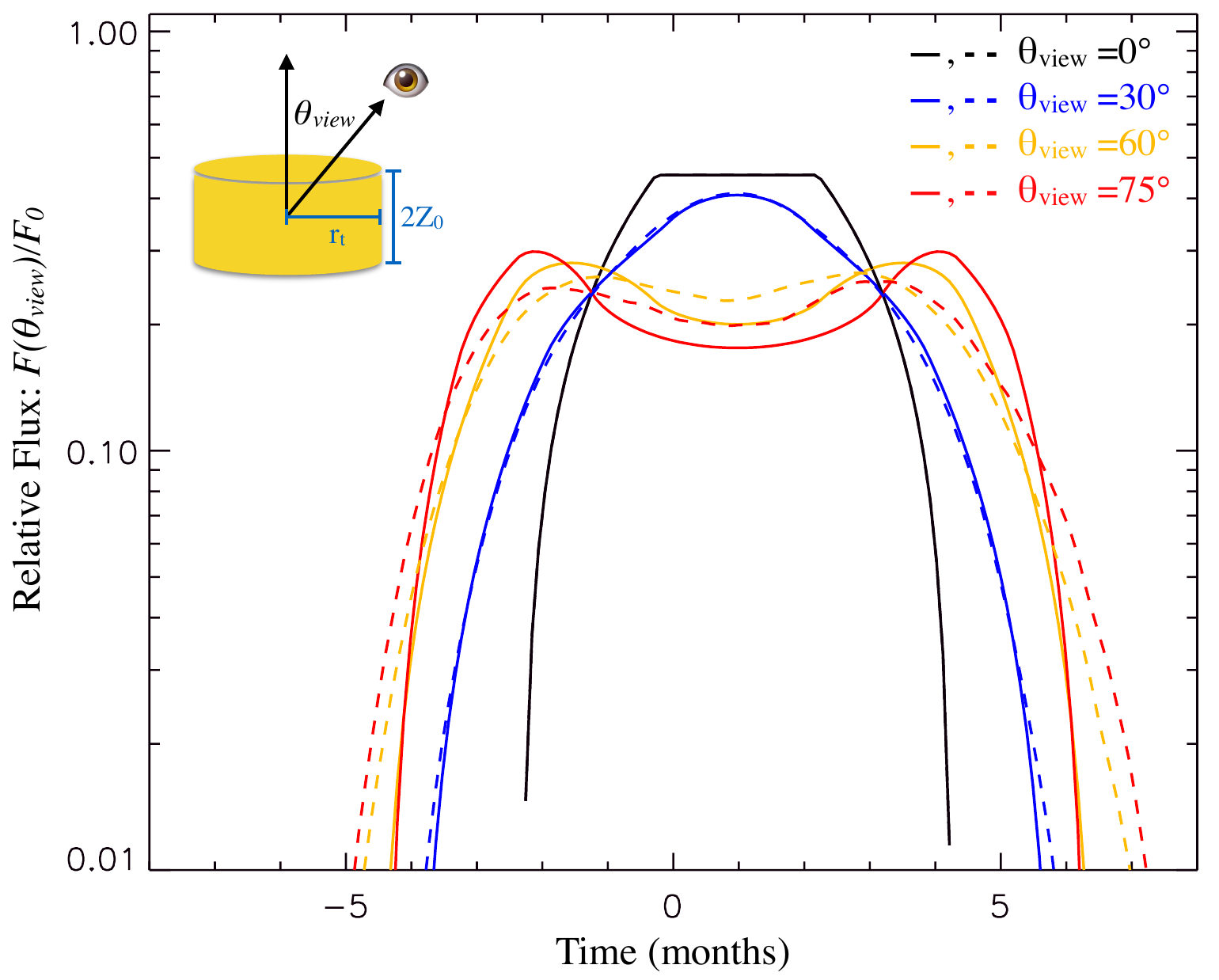}
 \caption{Relative flux viewed from different inclination angles considering the effect of geometrical light travel. The normalized flux $F_0$ is the flux treating the afterglow as a point source ($F_0 \equiv L/4\pi d^2_L$, where $d_L$ is the luminosity distance). The zero-point of time is defined as the afterglow's arising time if placed in the center. The solid and dashed lines indicate the cases of $r_t$ fixed at 0.1 pc and randomly distributed within 0.05--0.15 pc, respectively. The real flux is not so high as $L/4\pi d^2_L$, but should be lower. }  
 \label{fig7view} 
\end{figure}

If the inner edge of the torus is $\gtrsim 10^{-2}$ pc, the difference in the arrival time of photons from different emission zones can be several months, which is comparable to the duration the afterglows. This will smear the light curves and change predicted peak luminosity significantly. 
In order to explore the effect of geometrical light travel, we simplify the radiation zone as a cylindrical surface with a radius of $r_t$ and a height of $2Z_{0}$. We also explore a non-axisymmetric emission structure, in which the radial distances of emission zones are randomly distributed within 0.05--0.15 pc instead of being a fixed value of $r_t$. The duration of the afterglows is assumed to be $t_{\rm afg}=2$ months. The light curve observed from different inclination angles ($0^{\circ}$ denotes "face-on") is plotted in Figure \ref{fig7view} (see Figure \ref{figL1} for more cases).  
We find that, the light curve viewed from a large inclination angle is smeared significantly, and the flux is several times lower than that of an isotropic point radiation source. 
The reduction of the flux also depends on the relationship between $r_t/c$ and $t_{\rm afg}$. When $r_t/c < t_{\rm afg}$, the flux approaches the point source case, while $r_t/c \gtrsim t_{\rm afg}$, it is reduced significantly (Figure \ref{figL1}). 

Combined with our previous work (\citealt{mou2021}), we believe that the wind-torus interactions, or more generally, wind-cloud (dusty or not dusty) interactions, lead to simultaneously arising of X-ray, 
 radio-infrared emissions and gamma-rays, with time lags of the order of year.  
 Multi-band radiations provide a cross-validation for testing different physical models. 
The upcoming radio/X-ray surveys will expand the sample of TDEs, so that we can identify which transient sources may be the afterglows from TDE wind-torus interactions.

%%%%%%%%%%%%%%%%%%
%%%%%%%%%%%%%%%%%%

\section*{Acknowledgements}
 We are grateful to the anonymous referee for his/her careful reading, and providing valuable and in-depth comments that greatly improved the manuscript. 
We thank Zhuo Li and Ruoyu Liu for helpful discussions.  
G.M. is supported by National Science Foundation of China (NSFC-11703022, 11833007), ``the Fundamental Research Funds for the Central Universities'' (No. 2042019kf0040) and the opening fund of the Key Laboratory of Galaxy Cosmology, Chinese Academy of Sciences (No. 18010202). 
W.W. is supported by the National Program on Key Research and Development Project (Grants No. 2016YFA0400803) and the NSFC (11622326 and U1838103). 

\section*{Data Availability}
The data underlying this article will be shared on reasonable request to the corresponding author.

%%%%%%%%%%%%%%%%%%%%%%%%%%%
%%%%%%%%%     APPENDIX    %%%%%%%%%%
%%%%%%%%%%%%%%%%%%%%%%%%%%%

\appendix

\section{Physics of Wind Cloud Interactions}
\label{wdcld}
The velocity of the cloud shock is $\vsc \simeq\chi^{-0.5} \vw$, where $\chi \equiv \rhoc/\rhow$ is the density contrast. 
Here we refer the expression in \citet{mckee1975}. However, we must note that the expression in \citet{mckee1975} has different meaning in which it describes the interaction of post-shock ISM (rather than the supersonic wind) and cloud buried in it. 
For wind-cloud interactions we concern here, as a non-strict mathematical proof, during the interactions, the post-shock wind and the post-shock cloud should be in the pressure balance, otherwise the disturbances will travel quickly at the sound speed which is very high for post-shock materials. This is also verified by our hydrodynamic tests. 
The pressure of post-shock wind is $P_b\simeq\frac{3}{4}\rhow \vw^2$, and the pressure of post-shock cloud is $P_c\simeq \frac{3}{4}\rhoc \vsc^2$. Thus, we obtain the above expression which describes the interaction of supersonic wind and cloud.

Now, let's estimate the total energy driven into the cloud by wind.  
Within $dt$, the sum of kinetic energy and internal energy gained by cloud with covering factor of $\cv$ at $r$ is:  
\be
\begin{split}
dt \cv 4\pi r^2 \cdot \left[0.5\rhoc \vsc \cdot (0.75\vsc)^2+\rhoc \vsc (\gamma_{\rm ad}-1)^{-1} 3 \vsc^2/16\right] \\
=dt \cv 4\pi r^2 \cdot (9\rhoc\vsc^3/16+9\rhoc \vsc^3/16)= dt \cv 4\pi r^2 \cdot 9\rhoc \vsc^3/8,
\end{split}
\ee
where $\gamma_{\rm ad}=5/3$ is the adiabatic index, $0.75\vsc$ is the velocity of post-cloud-shock materials, and $(\gamma_{\rm ad}-1)^{-1} 3 \vsc^2/16$ is the internal energy per unit post-cloud-shock mass. Considering that during $dt$, the kinetic energy of the passing wind is $4\pi r^2 \vw dt \times 0.5\rhow \vw^2$, the ratio of the total energy driven into the cloud to the wind energy is $\simeq \cv \chi^{-0.5}$.

\section{Adiabatic Cooling of Cosmic Rays at Bow Shocks}
\label{adb}
\begin{figure*}
\centering
\includegraphics[width=0.92\textwidth]{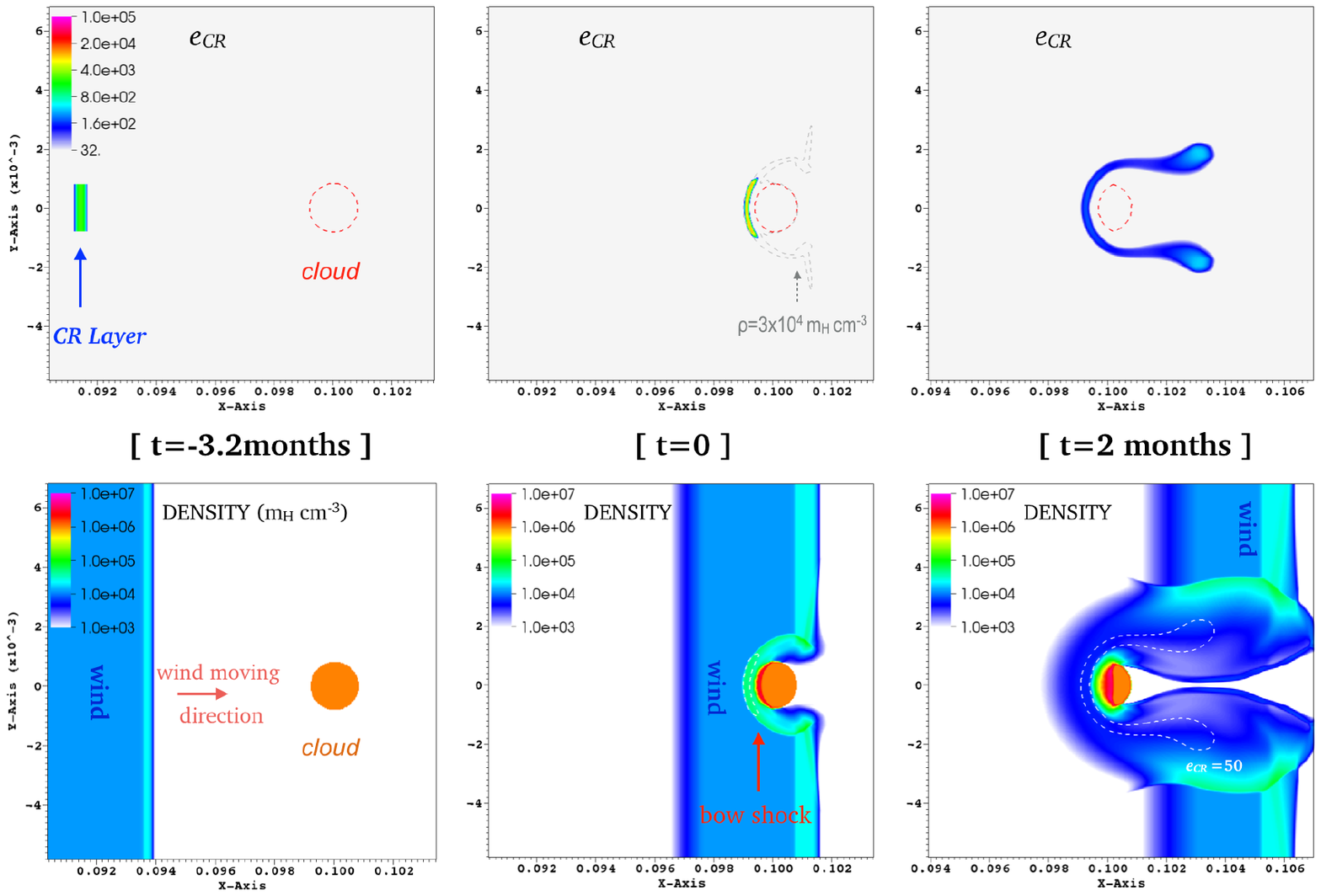}
 \caption{ Evolutions of distribution of CR's energy density (upper panels, in arbitrary units) and thermal gas density (bottom panels) for the model in which the duration of TDE wind is set to be 2 months. Coordinates are in units of parsec. 
Time \emph{zero} is set to be the moment when the CR's energy reaches its peak (CR layer just passed the bow shock). At time \emph{zero}, it can be regarded as a certain amount of CRs have been ``generated'' at the bow shock. 
Obviously, CRs do not evolve in a free-expanding way, but are squeezed by the shocked wind. Thus, the adiabatic cooling is not so fast as in the free-expanding case.  } 
\label{figB1}
\end{figure*}

The adiabatic cooling process of cosmic rays at the bow shock intrinsically is a two-fluid problem, of which CR can be treated as the second-fluid. This process is complex, and there may be a large error if the timescale of the adiabatic process of CR is approximated to 
the dynamical timescale (defined as the cloud radius over the shock speed).  
Here, we resort to hydrodynamic simulations to explore the adiabatic cooling process of cosmic rays at the bow shock, and choose ZEUS3D code (\citealt{clarke1996, clarke2010}) which can deal with two-fluid problems (one-fluid is the thermal gas with an adiabatic index of $\gamma_1=5/3$, while the other one is CR with an adiabatic index of $\gamma_2=4/3$). 
We run a series of tests, in which the CRs are injected with TDE wind in the form of a thin layer (see the top left panel in Figure \ref{figB1}). When the CR is swept up by the bow shock (the CR's energy will jump dramatically due to strong compression of the shock), it can be roughly regarded that the \emph{tracer CR} has been produced at that time. 
From this moment on, hydrodynamic simulations reveal the adiabatic process of the \emph{tracer CR}. 

The simulation domain is 0.028 pc $\times $ 0.016 pc in 2D Cartesian coordinates, which is divided into $1200\times 600$ meshes. The density of the cloud is $1\times 10^6 ~\mhcm$, and the wind density is $1\times 10^4 ~\mhcm$. The duration of the wind is 
 $t_{\rm bst}$, 
and the velocity is fixed at $3\times10^4~\kms$. At the moment when the wind has been injected halfway, we start to inject the CR with the same width as the cloud, and injection lasts for 3 days. In this way, we obtain a thin CR layer moving together with the wind (see Figure \ref{figB1}), which is in pressure balance with the ambient. 
After passing the bow shock, the energy density of CR is still much lower than that of thermal gas, thus it has negligible effect on the kinematics of thermal gas. Besides, we did not consider either the magnetic field or the diffusion of CRs.  

\begin{figure*}
 \centering
 \includegraphics[width=0.95\textwidth]{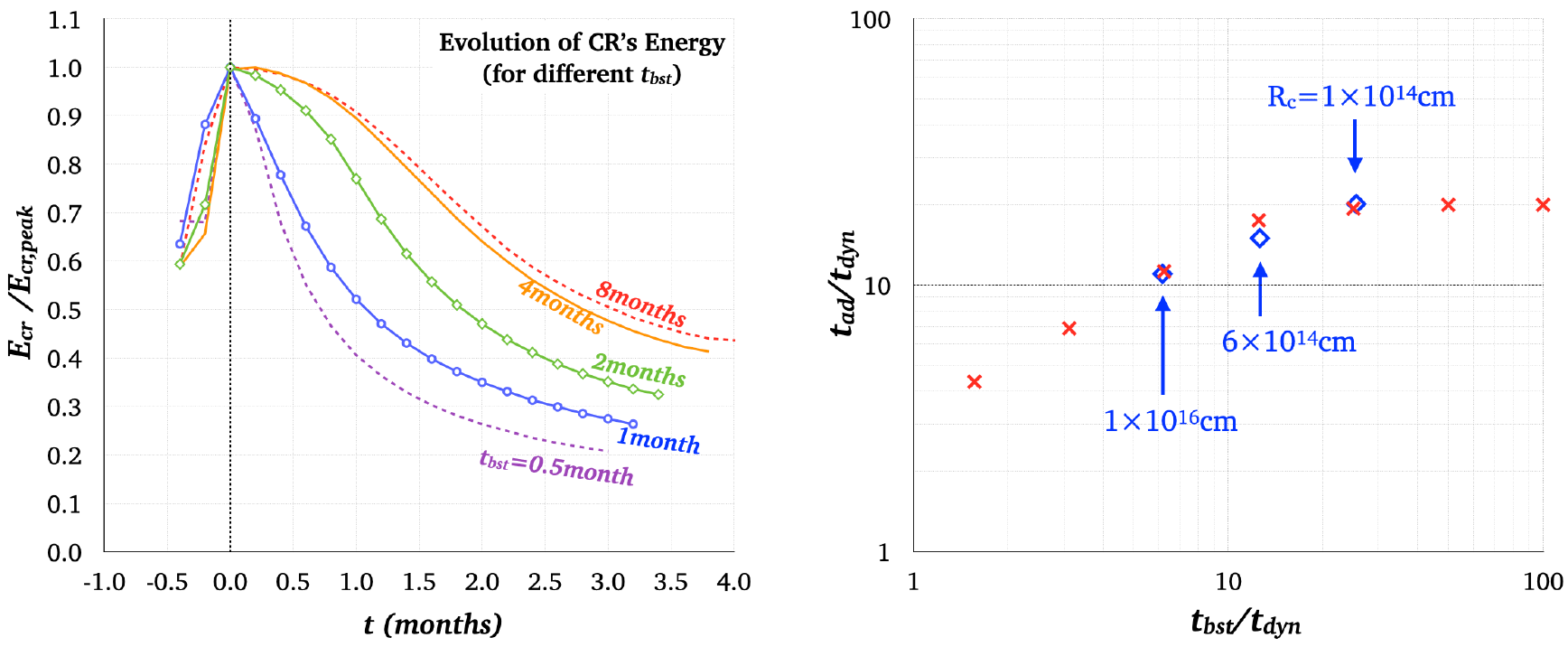}
 \caption{Left panel: evolution of CR's energy
  at cloud shocks 
 (normalized to the peak CR energy) for different $t_{\rm bst}$. The cloud radii are fixed at $\Rc=2.5\times 10^{15}$cm. 
Right panel: relationship between the adiabatic timescale (normalized to $\tdyn$) and the duration of wind (normalized to $\tdyn$). The adiabatic timescale $t_{\rm ad}$ is calculated by 2$\times t_{1/2}$ of which $t_{1/2}$ is the timescale declining from the peak energy value  to one half. Red crosses mark the result of $\Rc=2.5\times 10^{15}$cm, while blue diamonds mark the results of $\Rc=1\times 10^{16}$cm, $6\times 10^{14}$cm and $1\times 10^{14}$cm. When $t_{\rm bst} \gg \tdyn$, $t_{\rm ad}$ approaches $\sim 20 \times \tdyn$. Moreover, we find that this law applies to different cloud radii from $10^{14}$ cm to $10^{16}$ cm.  }   
\label{figB2}
\end{figure*}  

We plot the snapshots of CR-layer in Figure \ref{figB1}. CRs do not expand freely during the wind-cloud interacting stage or the subsequent stage, but are significantly confined. This will slow down the adiabatic loss of CR's energy compared with a free-expanding case. As shown in Figure \ref{figB2}, when the wind duration $t_{\rm bst}$ is significantly larger than the dynamical timescale $\tdyn$ ($\tdyn \equiv \Rc/\vw$), CR's adiabatic loss spends a much longer time than the dynamical timescale: $t_{\rm ad}=\kbow \tdyn \simeq 20\tdyn$. 
Moreover, in the limit $t_{\rm bst} \lesssim \tdyn$, adiabatic cooling of CR approaches the free-expanding process. In the main text, we adopt a conservative value of $\kbow=10$. 

\section{Adiabatic Cooling of Cosmic Rays at Cloud Shocks}
\label{adbcld}
\begin{figure}
    \includegraphics[width=1.0\columnwidth]{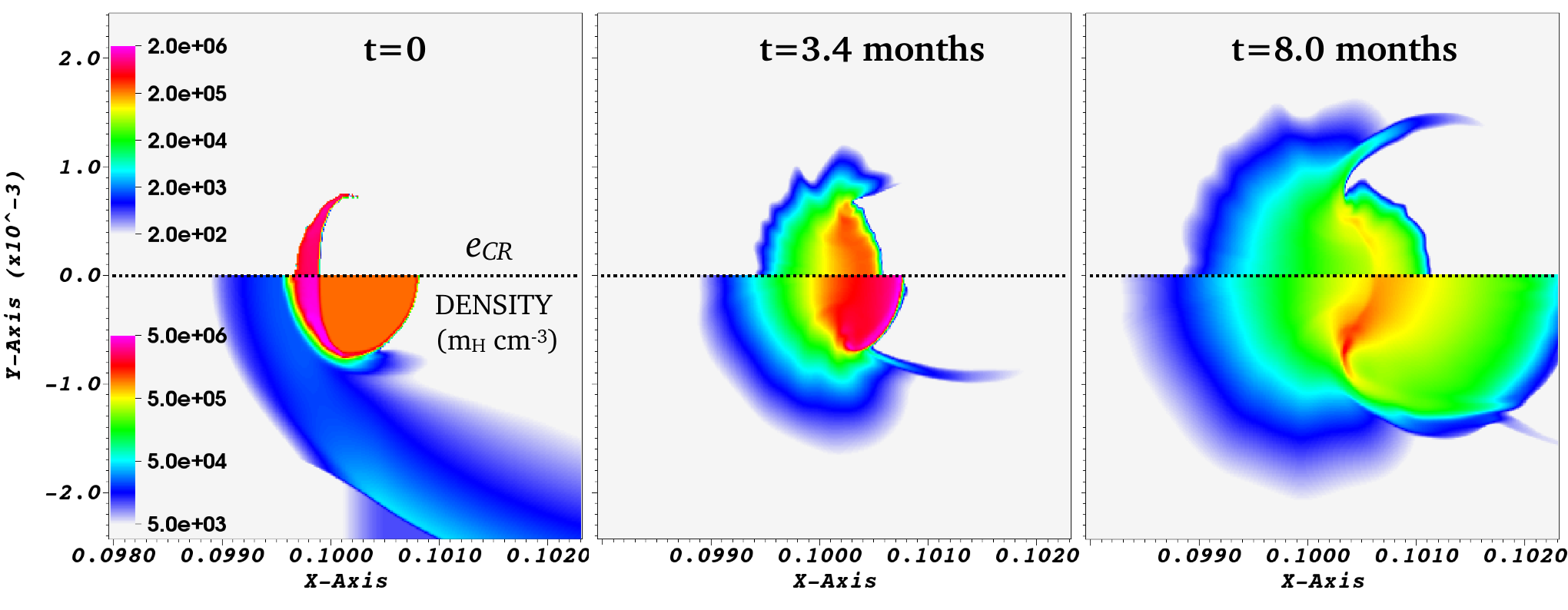}
    \includegraphics[width=0.9\columnwidth]{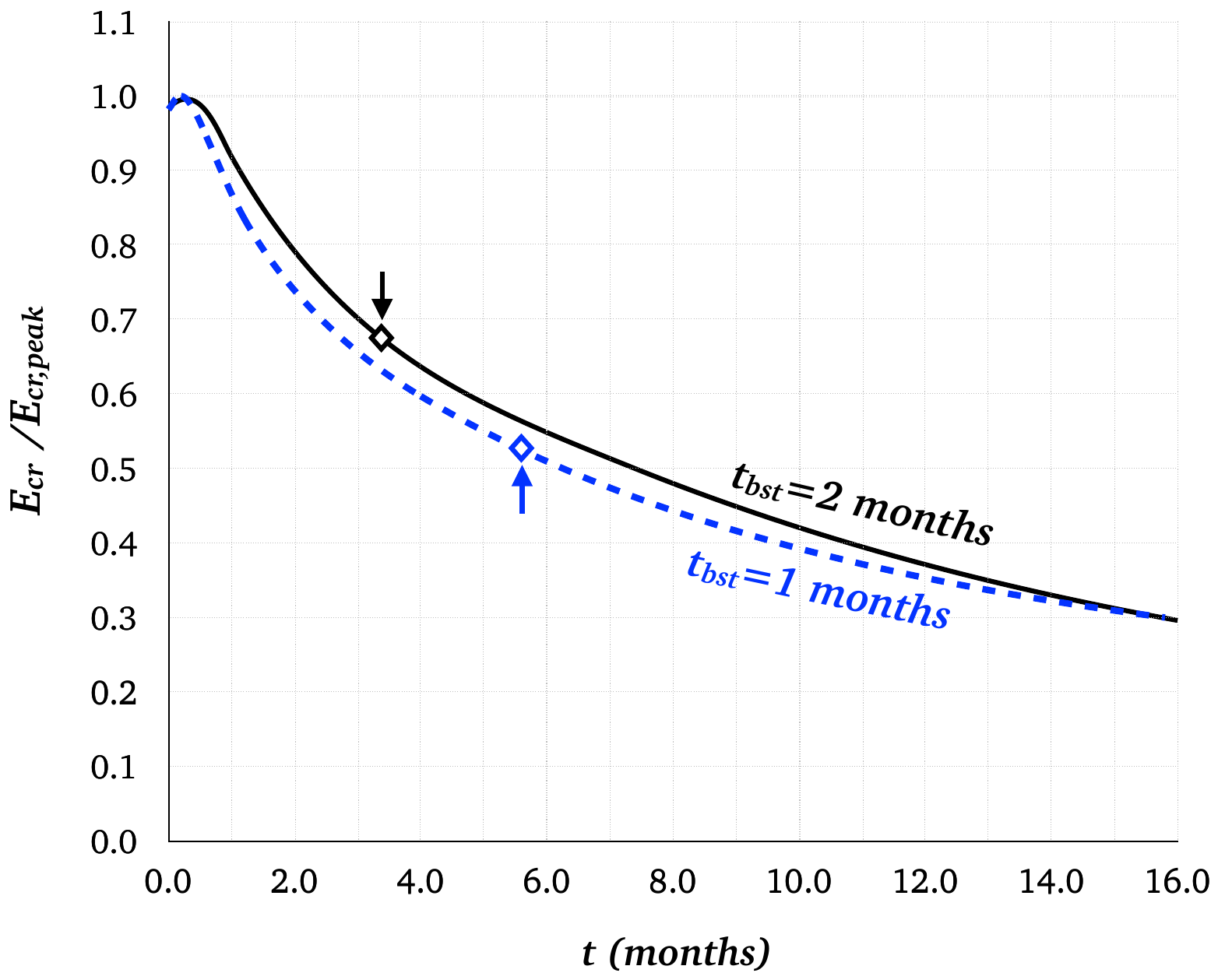}
 \caption{The upper panels show the evolution of CR's energy (at cloud shocks) and density for $t_{\rm bst}=2$ months.  The zero-point of time is defined as the moment when the tail end of the TDE wind is passing the cloud.
The lower panel shows the evolution of CR's total energy (at cloud shocks). The diamonds mark the moments when the cloud shock sweeps the entire cloud. }   
\label{figAC}
\end{figure}

Another question is whether the adiabatic cooling of CRs at the cloud shock is efficient. 
In our previous work (Appendix A.1 in \citealt{mou2021}), we found an interesting result that the internal energy of post-shock cloud remains almost a constant before the entire cloud has been swept over. This result also implies the answer to the above question: adiabatic loss of CRs at cloud shocks is not efficient before the cloud shock swept up the entire cloud. In order to check this, following appendix \ref{adb}, we preformed another two simulations. The cloud radius is fixed at $\Rc=2.5\times 10^{15}$cm. At the moment when the tail end of the TDE wind is passing the cloud, CRs are put in the grids behind the shock where the density compression ratio is higher than 2.0, and the energy density is set to be $e_{\rm cr} \propto \rhoc$. The snapshots of the CR energy are presented in Figure \ref{figAC}. We find that adiabatic loss of CR's energy is not quite efficient even after the cloud shock has swept the entire cloud. The loss of CR's energy is lower than $50\%$ within 6 months.  

\section{Gamma-rays and Neutrinos from pp Collisions} 
\label{ppc}

Collisions between CRp and thermal protons are able to produce pions if the CRp energy exceeds the threshold value of $E_p\simeq 1.4$ GeV. The reaction channels of pp collisions are: 
 \begin{gather}
 p+p \rightarrow p+p+a \pi^{0}+b(\pi^{+}+\pi^{-}), \\
 p+p \rightarrow p+n+\pi^{+}+a\pi^{0}+b(\pi^{+}+\pi^{-}),
 \end{gather}
where $a$ is generally equal to $b$ and they denote the number of pions produced in the reaction. 
The pions are very short-lived, and will instantly decay:
\begin{gather}
 \pi^{0} \rightarrow 2\gamma ,\\
 \pi^{+} \rightarrow \mu^{+}+  \num,~~ \mu^{+} \rightarrow e^{+} + \nue+\anum, \label{pi1} \\
 \pi^{-} \rightarrow \mu^{-}+\anum,~~ \mu^{-} \rightarrow e^{-}+\anue +\num. \label{pi2}
 \end{gather}
The decay of neutral pions induces a lower limit of $\Egm \sim70$ MeV for the gamma-rays, which is a characteristic signature for the pion-decay as in Fermi bubbles (\citealt{su2010, crocker2011, mou2015}). The decay of charged pions generates secondary high-energy electrons/positrons, which can also provide gamma-rays by the inverse Compton scattering on soft photons and the bremsstrahlung process. 

For pp collisions with protons only, the number of the secondary stable particles (including gamma-rays, electrons/positrons, and neutrinos) produced per unit time can be calculated by the following formula:
\be
\frac{dN_f(E_f)}{dE_f}= \int_{T_p} \frac{d\sigma(T_p, E_f)}{dE_f} v_p \nh \Np(T_p) dT_p , %\\
\ee
in which $f$ represents the species of secondary particles ($\gamma$, $\epm$, $\nue$, $\anue$, $\num$ and $\anum$), $\frac{d\sigma(T_p, E_f)}{dE_f}$ is the inclusive cross section as a function of both incident CRp's kinetic energy and the secondary particle's energy, $v_p \simeq c$ is the velocity of CRp, $\nh$ is the number density of thermal Hydrogen atom/nucleus, the expression of $\Np(T_p)$ is given by equation \ref{eqNp}, and $T_p\equiv E_p-m_p c^2$. The heavier elements (mainly Helium) will increase the above value by $\sim 50\%$ (\citealt{mori1997}). 

In a volume V, the total number of the secondary stable particles generation rate is:
\be
\begin{split} 
\frac{dN_f(E_f)}{dE_f}\simeq 1.5 \times \int_{T_p} \frac{d\sigma(T_p, E_f)}{dE_f} c \nh \Np(E_p) dT_p \\
= 1.5c \nh K_p(t) \cdot  \int_{T_p} \frac{d\sigma(T_p, E_f)}{dE_f} E^{-\Gamma}_p \exp(-E_p/\Emax) dE_p,  \\
\end{split}
\ee
in which we have added the correction by the Helium for the coefficient 1.5. The calculation formula is then divided into a term $\nh K_p(t)$ and a normalized reaction term 
 $\int_{T_p} \frac{d\sigma(T_p, E_f)}{dE_f} E^{-\Gamma}_p \exp(-E_p/\Emax) dT_p$. 
We calculate the normalized reaction term by cparamlib package \footnote{https://github.com/niklask/cparamlib} (\citealt{kamae2006, karlsson2008}).

\section{Gamma-rays from Inverse Compton Scatterings}
\label{ics}
The number of gamma-ray photons produced per unit time per unit energy from ICS of CRe, is given by:
 \be
 \begin{split}
  \frac{dN_{\gamma}(\Egm)}{d\Egm}= c \int_{E_e} \int_{E_{ph}}  \frac{d\sigma_{\rm IC}(\Egm, E_e, E_{ph})}{d\Egm} \Ne dE_{e} \frac{dn_{ph}}{dE_{ph}} dE_{ph}  \\
  \end{split}
  \label{icsdnde}
 \ee
where $\Ne$ is the energy distribution of primary CRe 
(see equation \ref{eqNe1} and \ref{eqNe2}). 
The differential cross sections of IC scattering is given by (\citealt{blumenthal1970}):
\be
 \begin{split}
\frac{d\sigma_{\rm IC}(\Egm, E_e, E_{ph})}{d\Egm}=\frac{3\sigma_{T}}{E_e \Gamma_{\epsilon}} \left[2q \ln q+(1+2q)(1-q)+ \frac{(\Gamma_{\epsilon} q)^2 (1-q)}{2(1+\Gamma_{\epsilon}q)} \right]
\end{split}
\ee
where $\sigma_T$ is the Thomson cross section and 
\be
\Gamma_{\epsilon} = \frac{4 E_{ph} E_{e}}{m^2_e c^4}, ~~~~ q=\frac{\Egm}{\Gamma_{\epsilon} (E_e-\Egm)}.
\ee

 \section{Synchrotron of CRe}
 \label{syn}
 A relativistic electron with an energy of $\gamma m_e c^2$ in a magnetic field of B will generate synchrotron emission:
 \be
 P_{\nu}(\nu, \alpha)=\frac{\sqrt{3}e^3 B \sin \alpha}{m_e c^2} \frac{\nu}{\nuc} \int^{\infty}_{\nu/\nuc} K_{5/3}(t)dt,
 \ee
where $P_{\nu}d\nu$ is the radiation power at $\nu \sim \nu+d\nu$, $\alpha$ is the pitch angle between electron velocity and the magnetic field (assuming that the pitch angle is random in this work), $\nuc$ is the critical frequency $\nuc = 3eB\gamma^2 \sin \alpha/(4\pi m_e c)$, and $K_{5/3}(t)$ is the modified Bessel function. 
Considering the probability distribution of the pitch angle at $\alpha \sim \alpha+d\alpha$ is $0.5\sin \alpha d\alpha$, we have the total synchrotron radiation power per unit frequency (in erg s$^{-1}$ Hz$^{-1}$) for a given distribution $\Ne(E_e)$ of CRe: 
\be
j_{\nu}= \int_{E_e} \int_{\alpha} P_{\nu}(\nu, \alpha) ~ \frac{1}{2} \sin \alpha ~d\alpha ~\Ne(E_e) dE_e . \\
\ee

\section{Gamma-Rays from Bremsstrahlung of CRe}
\label{bremcre}
The bremsstrahlung of a relativistic electron colliding with a charged static nucleus also contributes to the gamma-ray emissions.
In extreme relativistic case ($\gamma_{e} \gg 1$), the differential cross section of the bremsstrahlung is (e.g., \citealt{heitler1954}): 
\be
\frac{d\sigma_{br}(E, \Egm)}{d\Egm} d\Egm = \frac{4\alpha r^2_{0} Z^2}{\Egm}F(E, \Egm) d\Egm  ~,
\ee
in which $E$ is the energy of the relativistic electron, $E_{\gamma}$ is the photon energy, $\alpha=1/137$ is the fine structure constant, $r_0=e^2/m_e c^2=2.818\times 10^{-13}$ cm is the classical electron radius, and $Z$ is the number of nuclear charge. The function $F(E,E_{\gamma})$ depends on the the screening of the nucleus, and for a bare nucleus, it is 
\be
F(E,E_{\gamma})=\left[ 1+\left(1-\frac{E_\gamma}{E}\right)^2-\frac{2}{3} \left(1-\frac{E_\gamma}{E} \right) \right] \times
 \left[ \ln \left(\frac{2E(E-E_\gamma)}{m_e c^2 E_\gamma} \right)-\frac{1}{2} \right] .
\ee 
Considering that the shocked materials are heated to a high temperature ($\gtrsim 10^7$K in the cloud), the most majority of the atoms are fully ionized. Thus, we can use the approximate of the bare nucleus. 
The integral of $E^{-1}_{\gamma} F(E,E_{\gamma})$ over $E_{\gamma}$ from 0 to E is $\sim 10$. Hence, for a CRe colliding with hydrogen nuclei ($Z=1$), the total cross section is $\sim 20-30$mb.  
The cooling timescale due to bremsstrahlung is $t_{\rm br} \simeq 5{\rm yr} ~ n^{-1}_7$, which is much longer than ICS and adiabatic loss timescale. Therefore, the contribution of bremsstrahlung of CRe to the gamma-rays is negligible. 

\section{Contributions of Secondary Leptons from pp Collisions}
\label{2ndcre}
There are two main sources of CRe: one is from shock acceleration (primary CRe), and the other is the products $e^{-}e^{+}$ from pp collisions (so-called ``secondary CRe'').  

The secondary CRe undergo cooling via ICS and synchrotron radiation. The evolution of the energy distribution of secondary CRe can be obtained by equation \ref{evolsed}, in which the source term here is $S(E,t)=dN_e(E_e)/dE_e$ (equation D7, the subscript $e$ represents both electron and positron).  
The generation rate of secondary CRe limits its gamma-ray radiation. From Figure \ref{fig4sed}, the generation rate of $e^{\pm}$ is comparable to the gamma-rays from $\pi^{0}$ decays. This put an upper limit on the gamma-ray luminosity from the ICS of secondary CRe, which is in the order of $10^{39}\ergs$. Thus, we can ignore the ICS of secondary leptons.

\section{Photomeson Productions in p$\gamma$ Reactions}
\label{pgm}
In the environment of the photon number density exceeding gas number density, photomeson productions (p$\gamma \rightarrow N+K\pi$) may take place if the photon energy in the rest frame of the CRp is higher than 145 MeV. However, even for the most energetic CRp accelerated at bow shocks, the energy is $\sim 10^3$ TeV. Thus, the minimum energy of the seed photon required for this reaction is $\sim 0.1$ keV. For the radiation field at the inner edge of the torus ($\sim 10^{-3}~\ergcm$), assuming that the energy fraction of X-ray is $\sim 10\%$ percent, the X-ray photon density is $n_{\gamma} \sim 10^5~\cmc$. The timescale for photomeson production is $\tau_{p\gamma}=(\sigma_{p\gamma}cn_{\gamma})^{-1}\sim 10^5$ yr, where $\sigma_{p\gamma}\sim 10^{-28}$ cm$^2$ is the cross section (e.g., \citealt{kelner2008}). The reaction timescale of CR protons with energy lower than 1 PeV will be larger than this value due to fewer X-ray photons of higher energies. 
Therefore, we conclude that the photomeson production is negligible.

\section{Gamma-Ray Attenuation by Photon-Photon Pair Production} 
\label{gmgm}

\begin{figure}
   \includegraphics[width=1.0\columnwidth]{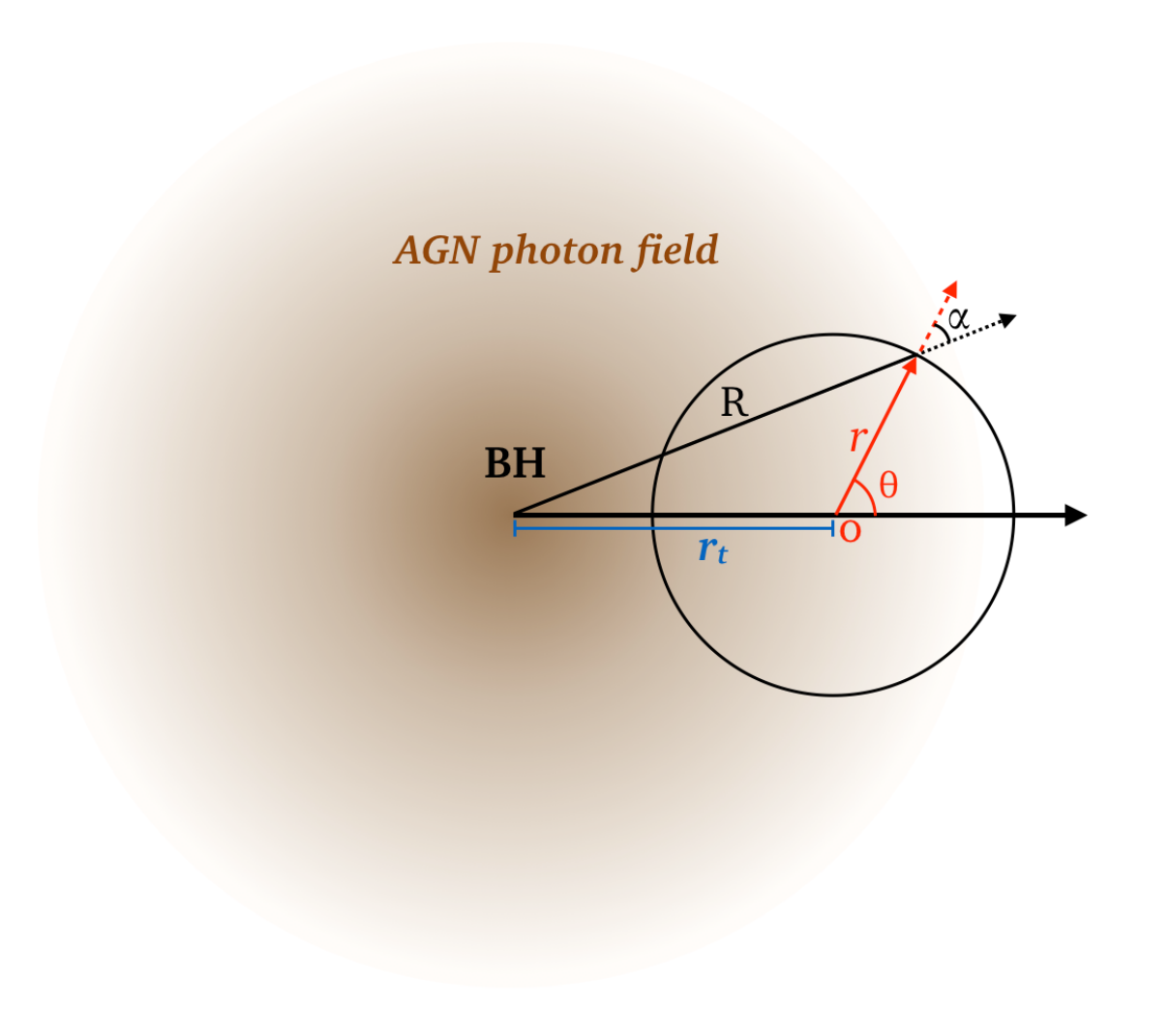}
   \caption{ The coordinate system adopted to analyze the escape probability of one gamma-ray in AGN photon field. The origin of the coordinates is set to be the generation position of the gamma-ray, and the emergent angle $\theta$ is the angle between the vector from the BH to the origin point and the gamma-ray's momentum. The photon-photon collision angle is marked as $\alpha$. $r_t$ is the distance from the inner edge of the torus to the BH, which is set to be 0.1 pc here.  }
 \label{figJ1} 
\end{figure}

The cross section for photon-photon reaction ($\gamma \gamma \rightarrow e^{\pm}$) is (e.g., \citealt{inoue2019}):
\be
\sigma_{\gamma\gamma}(\Egm, \nu, \alpha)=\frac{3\sigma_T}{16}(1-\beta^2) \left[2\beta(\beta^2-2)+(3-\beta^4) \ln\frac{1+\beta}{1-\beta}  \right] ,
\ee
where $\beta=\sqrt{1-\frac{2m^2_e c^4}{h\nu \Egm(1-\cos \alpha)}}$, and $\alpha$ is the collision angle of two photons in laboratory frame.  % take the line connecting the BH to the origin as the polar axis
In order to investigate the escape probability of a gamma-ray photon, it is convenient to establish a spherical coordinate system as Figure \ref{figJ1}. Then we have $\cos \alpha=(r^2+R^2-r^2_t)/(2rR)$, where $R^2=r^2+r^2_t+2rr_t \cos\theta$. 
For simplicity, here we assume that the gamma-ray's momentum direction is random. Considering that the probability of an emergent angle between $\theta$ and $\theta+d\theta$ is  $\frac{1}{2}\sin\theta d\theta$, the angle-averaged optical depth for the gamma-ray photon can be calculated by 
\be
\tau_{\gamma\gamma}=\int_{\theta} \int_{r} \int_{\nu} \sigma_{\gamma\gamma}(\Egm, \nu, \alpha) n_{\rm ph}(\nu, r, \theta) 
\frac{\sin\theta}{2} d\nu dr d\theta , 
\ee
where the number density of AGN photon per unit frequency is given by: 
\be
n_{\rm ph}(\nu, r,\theta) =\frac{L_{\nu, \rm AGN}  }{4\pi R^2 c \nu} .
\ee
The optical depth $\tau_{\gamma\gamma}$ is plotted 
{\rm in the upper panel in}
Figure \ref{fig4sed}. We find that the absorbed gamma-rays in pair production are mainly high-energy components of $\gtrsim 60$ GeV, and the absorbed gamma-ray energy 
 is $8\times 10^{40}~\ergs$, which 
will be converted into relativistic electron-positron pairs, and gamma-rays in lower energies via ICS on the AGN photons. However, this should not significantly alter the final gamma-ray spectrum.  

\section{Simulations on Effect of TDE wind - BLR Interactions} 
\label{blr}

\begin{figure}
   \includegraphics[width=0.95\columnwidth]{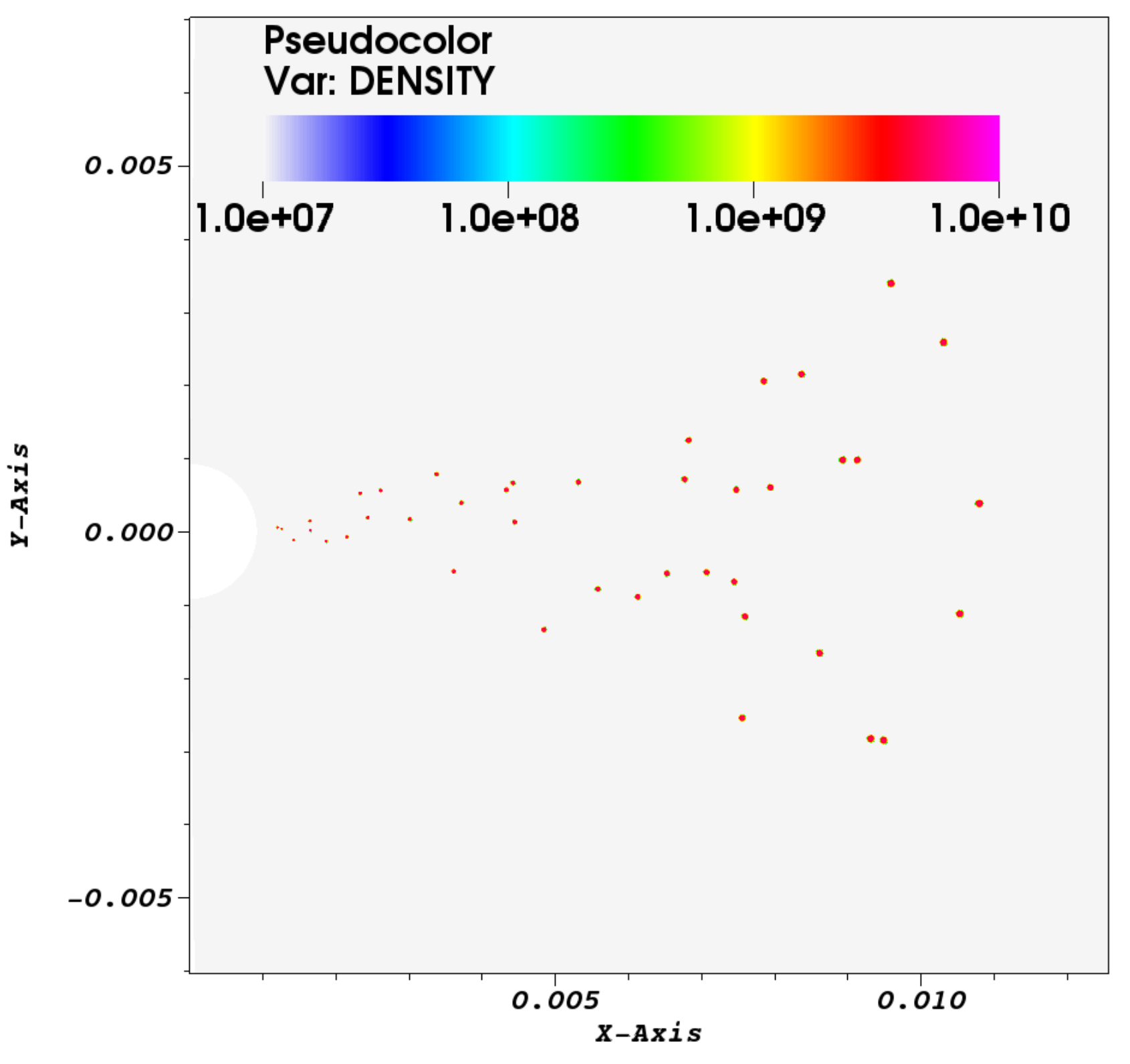}
   \includegraphics[width=0.95\columnwidth]{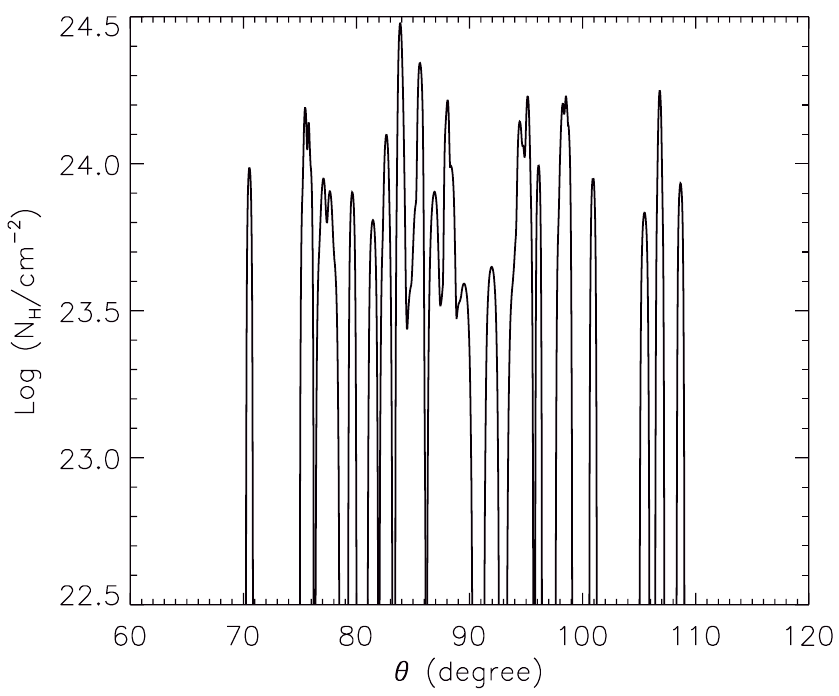}
   \caption{Upper panel: the initial setup of BLR that consisting of 41 small clouds. Lower panel: the column density of BLR measured from different $\theta$-angles. }
 \label{figK1} 
\end{figure}

 The effect of the TDE wind-BLR interactions is explored by hydrodynamic simulations with ZEUSMP code (\citealt{hayes2006}).  We choose 2.5D spherical coordinates and assume that $\partial_{\phi} =0$ where $\phi$ is the rotational direction of BLR. 
One difficulty is how to embody the BLR, and we simply assume that the BLR is made of small clouds distributed from $r=$ 0.001 pc to 0.011 pc, corresponding to a Keplerian velocity of 6000-2000 $\kms$ around a $10^7~ M_{\odot}$ BH. The clouds are randomly distributed within a wedge zone: $|Y|/(X-0.001{\rm pc}) < 0.4$ (see Figure \ref{figK1}). The cloud radius scales with the distance to the BH as $R_{\rm c,blr} = 4.6\times 10^{13}{\rm cm} \cdot (r/0.001{\rm pc})^\beta$ where $\beta=0.5$. The cloud density is set to be $5\times 10^9~{\rm m_H cm^{-3}}$. The covering factor of such a BLR is 0.18 for $N_H>10^{23} $ cm$^{-2}$. 
The simulation domain is $r \in [9.2\times 10^{-4} ~ {\rm pc}, 7.0\times 10^{-2} ~{\rm pc} ]$, and $\theta \in [0^{\circ},180^{\circ}]$, which is divided into 7200 pieces in $r-$direction ($dr_{i+1}/dr_i$ = 1.0006) and 4800 uniform pieces in $\theta-$direction.  
The TDE wind is injected at inner boundary with a velocity of 0.1c and a mass outflow rate of $5.7 ~\msunyr$. The duration of the TDE wind is 2 months. 

The X-ray emission from the wind-BLR interactions is largely dependent on the physics of BLR cloud, which remains quite uncertain currently. The ram pressure of the wind is around the order of $10^3$ dyn cm$^{-2}$ at $r\sim10^{-3}$ pc. 
Such a high ram pressure will squeeze the cloud into a very dense phase: $n=P_{\rm ram}/(k_B T)= 7\times 10^{14} ~ {\rm cm^{-3}} P_3 T_4$ ($P_3 \equiv P_{\rm ram}/10^3~{\rm dyn ~cm^{-2}}, T_4\equiv T/10^4 K$). The temperature of $10^4$ K is a characteristic value for a typically low photoionized gas around an AGN (the lower branch in the photoionization equilibrium curve), and it will be further lower for an optically thick cloud. Due to the high density of shocked clouds, the subsequent energy conversion efficiency (from wind to cloud) will significantly decrease. To simulate this process, we must not only resolve the initial $10^{14}$ cm-sized cloud, but also resolve $\sim 10^{9}$ cm-sized compressed clumps (the compression is mainly along radial direction), while the entire distribution region of BLR ($\sim 10^{16}$ cm-sized) should also be covered. This is technically difficult. 
On the other hand, if considering the possible magnetic field required for constraining the BLR clouds (\citealt{rees1987}), magnetic pressure may play a significant supporting role and prevent the cloud from further compression. In this case, the X-ray emission relies on the magnetic field strength and configuration. 
Moreover, the X-rays from post-shock materials would be (partially) absorbed by the clouds, and reprocessed in near-ultraviolet to optical band (\citealt{moriya2017}). Thus, taking into account these factors, here we did not explore the radiations from wind-BLR interactions.  

\section{Effect of Geometrical Light Travel Delay} 
\label{geom}
\begin{figure}
   \includegraphics[width=0.95\columnwidth]{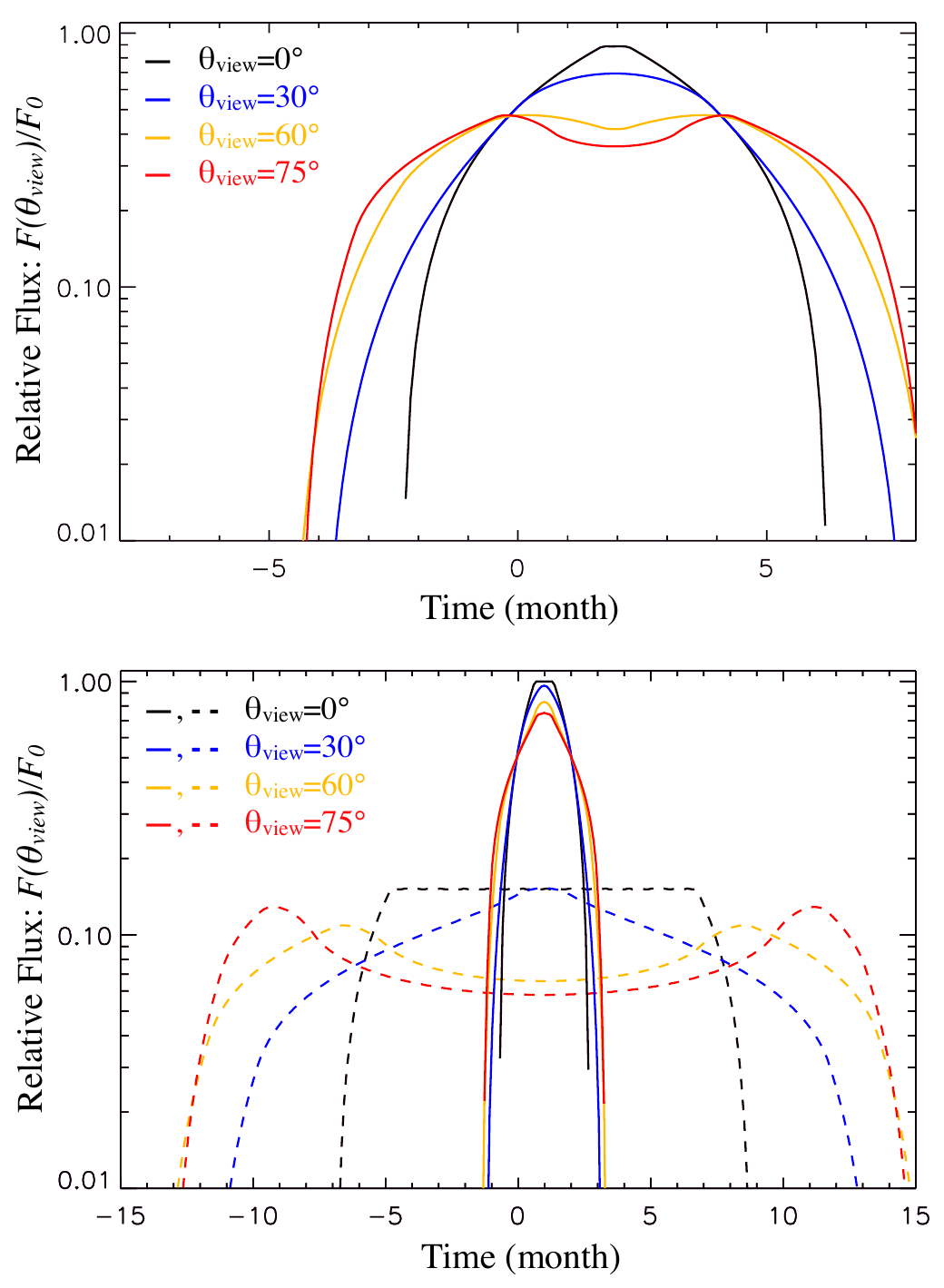}
   \caption{ In upper panel, we plot the relative fluxes from different viewing angles for $(r_t, Z_0, t_{\rm afg})$=(0.1 pc, 0.0577 pc, 4 months). In the lower panel, we plot the relative fluxes for $(r_t, Z_0, t_{\rm afg})$= (0.03 pc, 0.0173 pc, 2 months) in solid lines, and  (0.3 pc, 0.173 pc, 2 months) in dashed lines, respectively.  }
 \label{figL1} 
\end{figure}

We considered several cases to study the effect of light travel delay on the light curve and fluxe of the afterglows. 
In Figure \ref{figL1}, we plot the results of ($r_t$, $Z_0$, $t_{\rm afg}$)=(0.1 pc, 0.0577 pc, 4 months), (0.03 pc, 0.0173 pc, 2 months) and (0.3 pc, 0.173 pc, 2 months). 

%%%%%%%%%%%%%%%%%%%%%%%%%%%%%%%%%%%%%%%%%%%%%%%%%%

% Don't change these lines
\bsp	% typesetting comment
\label{lastpage}
\end{document}